\documentclass[10pt,twocolumn,twoside]{IEEEtran}

\usepackage{stfloats}
\usepackage{caption}
\usepackage{bm}   
\usepackage{amsmath}
\allowdisplaybreaks[4]
\usepackage{amssymb}
\usepackage{cuted}
\usepackage{indentfirst}  
\usepackage{graphicx}  
\usepackage{subfigure}
\usepackage{multirow} 
\usepackage{booktabs} 
\usepackage{cite}  
\usepackage{threeparttable}   
\usepackage{algorithm}
\usepackage{algorithmic}
\usepackage{arydshln}
\usepackage{color}
\usepackage[american]{babel}
\usepackage{microtype}
\usepackage[numbers,sort&compress]{natbib}  
\usepackage{steinmetz}
\usepackage{ifthen}
\usepackage{cases}
\usepackage{enumerate}
\usepackage{epstopdf}
\usepackage{float}
\usepackage{ulem}
\usepackage[cp1251]{inputenc}

\def\mV{\mathcal{V}}

\def\mE{\mathcal{E}}
\def\mU{\mathcal{U}}

\def\mbR{\mathbb{R}}
\def\mbP{\mathbb{P}}

\def\mbE{\mathbb{E}}
\def\mP{\mathcal{P}}

\begin{document}
\title{Stability Constrained OPF in Microgrids: \\ A Chance Constrained Optimization Framework with Non-Gaussian Uncertainty}

\author{Jun~Wang,~\IEEEmembership{Graduate Student Member,~IEEE,}
        Yue~Song,~\IEEEmembership{Member,~IEEE,}
        David~J.~Hill,~\IEEEmembership{Life Fellow,~IEEE,}
        Yunhe~Hou,~\IEEEmembership{Senior Member,~IEEE,}
        and Feilong~Fan,~\IEEEmembership{Member,~IEEE,}

\thanks{J. Wang, Y. Song, D. J. Hill and Y. Hou are with the Department of Electrical and Electronic Engineering, The University of Hong Kong, Hong Kong (e-mail: wangjun@eee.hku.hk; yuesong@eee.hku.hk; dhill@eee.hku.hk; yhhou@eee.hku.hk). F. Fan is with the College of Smart Energy, Shanghai Jiao Tong University, Shanghai, China (e-mail: feilongfan@sjtu.edu.cn).}
\vspace{-15pt}
}

\markboth{IEEE TRANSACTIONS ON xxx}  
{Wang \MakeLowercase{\textit{et al.}}: Stability Constrained OPF in Microgrids: A Chance Constrained Optimization Framework with Non-Gaussian Uncertainty}

\maketitle

\begin{abstract}
To figure out the stability issues brought by renewable energy sources (RES) with non-Gaussian uncertainties in isolated microgrids, this paper proposes a chance constrained stability constrained optimal power flow (CC-SC-OPF) model. Firstly, we propose a bi-level optimization problem, of which the upper level aims to minimize the expected generation cost without violating the stability chance constraint; the lower level concerns about the stability index given by a semi-definite program (SDP). Secondly, we apply the Gaussian mixture model (GMM) to handle the non-Gaussian RES uncertainties and introduce analytical sensitivity analysis to reformulate chance constraints with respect to stability index and operational variables into linear deterministic versions. By incorporating linearized constraints, the bi-level model can be efficiently solved by Benders decomposition-based approach. Thirdly, we design a supplementary corrective countermeasure to compensate the possible control error caused by the linear approximation. Simulation results on the 33-bus microgrid reveal that compared to benchmarking approaches, the proposed model converges 30 times faster with more accurate solutions.
\end{abstract}


%
\IEEEpeerreviewmaketitle
\vspace{-10pt}
\section{Introduction}
Towards a carbon neutral future, microgrids (MGs) as a critical component in future smart grids have been proposed and attracted much attention \cite{b1}. MGs with increasingly integrations of distributed renewable energy sources (RESs) can operate locally as single entities \cite{b2}. The energy schedule and operating strategy are always handled by solving optimal power flow (OPF) problems \cite{b3}. Due to the intermittency nature of RESs, OPF problems in MGs is more complex and uncertain \cite{b4}. To cope with the uncertainty brought by RESs in MGs, various methods have been investigated including three main types, robust optimization (RO) \cite{b5}, \cite{b6} stochastic optimization (SO) \cite{b7} and chance constrained optimization (CCO). RO is likely to provide a too conservative solution and SO may obtain improper decisions with constraint violations due to the discrepancy between the actual uncertainty and the representative scenarios in SO. CCO is another approach to address the impact of uncertainty in a less conservative way than RO and a more comprehensive way than SO \cite{b8}. According to the estimated PDF of uncertainties, CCO problems are mostly solved by converting chance constraints (CCs) into deterministic constraints \cite{b9}-\cite{b14}.

Among the existing literature, CCO in MGs mainly concerns about energy constraints. For instance, in \cite{b9}, two CCO problems are formulated, whose CCs include generation outputs and state of charge (SOC) of batteries. Inspired by \cite{b9}, \cite{b10} investigates distributionally robust CCO for islanded MGs, which additionally considers CCs for load demand and power balance. Later, steady-state security constraints, such as voltage constraint, line flow constraint, and post-contingency security, have been augmented into the CC-OPF problems \cite{b13}-\cite{b15}. Moreover, because of the natural complexity of CC, CC-OPF model mainly adopts the linear power flow or DC power flow, which influence the accuracy of solutions.

Actually, apart from steady-state security constraints, stability issue  cannot be negligible in MGs. The reasons are twofold. Firstly, the intermittent nature and fluctuated output of RES bring more disturbances in MGs \cite{b16}. Secondly, distributed generators (DGs) will decrease the system inertia since they are interfaced by inverters. Thus, inverter-interfaced MGs hold a smaller capacity to resist disturbances \cite{b17}. But, few CC-OPF models in existing works incorporate stability constraints. The stability evaluation usually requires accurate information of the operating equilibrium point which is described by AC power flow (ACPF) equation. However, once the CC-OPF problem adopts ACPF, it becomes nonconvex and NP hard, let alone the additional complexity brought by the stability CC. The most challenging part of transforming the stability CC is that the stability index hardly has explicit analytical sensitivity w.r.t. uncertain variables. A straightforward method to circumvent this obstacle is applying the Monte-Carlo or clustering method to obtain the numerical perturbation-based sensitivity \cite{b21}, which brings a heavy computation burden. This feature impedes the stability constraint being added in CC-OPF if the solution should be given within a relatively short timeslot (e.g. 15 minutes). To our knowledge, a systematic study of efficient stability CC reformulations has not been addressed.

Furthermore, in order to transform CCs into deterministic second-order cone constraints, which result in an optimization framework a second-order cone programming (SOCP), CCO mostly assumes that the uncertainty follows a Gaussian distribution. While, this assumption causes inaccuracy in modelling because RES uncertainties may follow non-parametric distributions \cite{b22}. In particular, non-Gaussian issue has not been well investigated in CC-OPF problems so far.

To fill the aforementioned research gaps both in the modeling and solution methodology, this paper establishes a chance constrained stability constrained OPF (CC-SC-OPF) scheme to ensure system stability under non-Gaussian uncertainties in isolated microgrids. Firstly, we design a bi-level optimization framework, in which the upper level problem focuses on minimizing the expected generation cost without violating stability CC; the lower level problem aims to calculate the stability index given by a semi-definite programming (SDP). Secondly, we apply the Gaussian mixture model (GMM) to well estimate non-parametric distributions of RES forecast errors. Through the estimated PDF, we utilize the analytical sensitivity analysis to transform CCs of stability index and operational variables into linear deterministic formulae. With the reformulated CCs, the bi-level model can be efficiently solved by Benders decomposition-based approach. We adopt the modified IEEE 33-bus system to validate the proposed CC-SC-OPF model. Simulation results reveal that compared to benchmarking approaches, the proposed method can converge much more quickly and achieve higher accuracy of the solution.

The major merits of the proposed method are twofold:

\begin{enumerate}
        \item We design a novel CC-OPF model which incorporates the stability CC. Moreover, compared to the conventional numerical perturbation for stability sensitivity evaluation, the proposed method obtains linear approximation of stability CC via analytical stability sensitivity which speeds up the computational efficiency by more than 30 times.

        \item The non-Gaussian uncertainty is efficiently handled by GMM with slight additional computation burden. These features add the practicality of CC-SC-OPF and facilitate its online implementation.
    \end{enumerate}

The rest of the paper is organized as follows. The dynamic model and stability analysis of microgrids are given in Section II. Section III proposes a bi-level CCO framework with considering stability issues. Section IV provides GMM-based non-Gaussian uncertainty estimation and CC reformulation approaches through the approximated sensitivity analysis. Benders based-algorithm and supplementary corrective measure for solving the proposed model is designed in Section V. A case study is given in Section VI to validate the effectiveness of the proposed CC-SC-OPF. Section V concludes this paper.

\vspace{-5pt}
\section{Dynamic Modelling}
\subsection{Modelling of Inverter-Interfaced Microgrid}
An islanded microgrid considered in this paper contains $n$ buses including $g$ dispatchable DG (hereinafter referred to as DG) buses, and  $d=n-g$ load buses. DG buses refer to buses connecting a DG and may connecting loads, while load buses refer to buses connecting loads and RESs. We denote DG buses as $\mV_G=\{1,2,...,g\}$, and load buses as $\mV_L=\{g+1,...,n\}$. In this paper, RESs are modelled as power injections with uncertainties, while DGs are interfaced by droop controlled inverters.
Then we formulate the dynamic model of the isolated microgrid. The ACPF is adopted in order to accurately characterize the equilibrium point for stability evaluation. Moreover, to avoid the existence of the zero eigenvalue in the Jacobian matrix, we set bus 1 as the angle reference to introduce relative angles $\alpha_{i}=\theta_{i}-\theta_{1}$, $i\in \mV_G \cup \mV_R \cup \mV_{L} \setminus \{1\}$, where $\theta_i$ denotes the node angle. Thus, the power flow takes the following form in \eqref{powerinj} w.r.t. $\alpha_{i}$.
\vspace{-5pt}
\begin{equation}\label{powerinj}
\begin{split}
    P_i^\textup{inj} &= {P_{Gi}} +{P_{Ri}} - {P_{Li}} \\
    &= V_i(\sum\limits_{j = 1}^nV_jG_{ij}\cos\alpha_{ij}+V_jB_{ij}\sin\alpha_{ij})\\
    Q_i^\textup{inj} &= {Q_{Gi}} +{Q_{Ri}} - {Q_{Li}} \\
    &= V_i(\sum\limits_{j = 1}^nV_jG_{ij}\sin\alpha_{ij}-V_jB_{ij}\cos\alpha_{ij})
\end{split}
\end{equation}
where $P_{Gi}$ and $Q_{Gi}$ denote the DG active and reactive power outputs; ${P_{Ri}}$, ${Q_{Ri}}$ and ${P_{Li}}$, ${Q_{Li}}$ represent the active and reactive power supply and demand of RESs and loads at the bus $i$, respectively; $V_i$ denotes the bus voltage magnitude; $G_{ij}$ and $B_{ij}$ are the $(i,j)$ entries in the conductance matrix $\bm{G}\in \mbR^{n\times n}$ and susceptance matrix $\bm{B}\in \mbR^{n\times n}$; $\alpha_{ij}$ denotes the difference between $\alpha_i$ and $\alpha_j$. Following the modeling process in our previous work \cite{b23}, by combining \eqref{powerinj} and DG droop dynamics, we formulate the following system-wide dynamic algebraic equations (DAE)
\begin{subequations}\label{dynamicmodel}
\begin{align}
   \dot{\alpha}_i &= \omega_{b}(\omega_i-\omega_1), i\in \mV_G\cup \mV_L\setminus \{1\} \label{angledynamics}\\
   {{\dot \omega }_i} &= f_{Gi},i\in \mV_G \label{Gfredynamics}\\
   {{\dot V }_i} &= h_{Gi},i\in \mV_G \label{Gvoldynamics}\\
   0 &= f_{Li}, i\in \mV_L \label{loadPdynamics} \\
   0 &= h_{Li}, i\in \mV_L \label{loadQdynamics}
\end{align}
\end{subequations}
in which the functions $f_{Gi}$, $h_{Gi}$, $f_{Li}$ and $h_{Li}$ are
\begin{subequations}
\begin{align}
   f_{Gi}&={K_{pi}}\!{F_{pi}}( -P_i^\textup{inj}\!+\!P_{Ri}\!-\!P_{Li} \!+\!P_{Gi}^*)-\! F_{pi}\!({\omega _i}\!-\!\omega^*)\\
   h_{Gi}&={K_{qi}}\!{F_{qi}}(\! -Q_i^\textup{inj}\!+\!Q_{Ri}\!-\!Q_{Li}+\!Q_{Gi}^*)\!-\! F_{qi}\!({V_i}\!-\!V^*_i)\\
   f_{Li}&=-P_i^\textup{inj}+P_{Ri}-P_{Li}\\
   h_{Li}&=-Q_i^\textup{inj}+Q_{Ri}-Q_{Li}
\end{align}
\end{subequations}
where $\omega_i, \omega_b$ denote angular frequency and the base frequency ($120\pi$ rad/s used in this paper), respectively; $\omega^*$, $V_i^*$, $P_{Gi}^*$ and $Q_{Gi}^*$ denote the set points for frequency, bus voltage, active and reactive generation; $K_{pi}$ and $K_{qi}$ denote the frequency and voltage droop gains of DG inverter;
$F_{pi}$ and $F_{qi}$ denote the corner frequencies of the first-order low-pass filters for measurement.

With considering the uncertainty of the RES outputs, we denote $\epsilon_{Ri}$ as the forecast error on the RES bus $i$, i.e. $P_{Ri}=\bar{P}_{Ri}+\epsilon_{Ri}$, where $\bar{P}_{Ri}$ is the forecast value. And we assume that the RES reactive power generation $Q_{Ri}$ follows the active power generation with a constant power ratio $\lambda$, such that $ Q_{Ri}=\lambda(\bar{P}_{Ri}+\epsilon_{Ri})$. The uncertain variable $\bm{\epsilon}_R=\{\epsilon_{Ri}\}$ has inherent dependence resulting from geographical connections among multiple RESs, and follows a non-Gaussian distribution. With RES uncertain forecast errors, the system power balance can be still achieved through an automatic power redispatch brought by DG droop control schemes.
In the small signal stability analysis, the forecast error has a significant impact on the equilibrium point, and may deteriorate the stability. To better underline the influence of RES uncertainties on the system stability and the countermeasures, we assume for simplicity that RESs and loads in this paper are all frequency independent and voltage independent in the proposed model. Also, the proposed model can extend to more general cases, such as frequency-dependent loads with few modifications.

\subsection{Small Signal Stability Index}
To generate the small signal model of the microgrid, according to system-wide DAE \eqref{dynamicmodel}, we linearize around the equilibrium point $(\bm{x},\bm{y})$, where $\bm{x}=[\bm{\alpha}_r^T, \bm{\omega}_G^T, \bm{V}_G^T]^T \in \mbR^{3g-1}$, $\bm{y}=[\bm{\alpha}_L^T, \bm{V}_L^T]^T\in \mbR^{2d}$ are state variable and algebraic variable, the variables with subscript ``G'' and ``L'' refer to the entries indexed by DG buses and load buses, respectively.
 In $\bm{x} $ and $\bm{y}$, $\bm\alpha_r = [\alpha_2, \alpha_3, ..., \alpha_g] \in \mbR^{g - 1}$, $\bm\omega_G$, $\bm V_G$ denote angular frequency and voltage magnitude at DG buses, respectively. And $\bm\alpha_{L} = [\alpha_{g+1}, \alpha_{g+2}, ..., \alpha_n] \in \mbR^d$, $\bm V_{L}$ are angle and voltage magnitude at load buses. What should be noted that since we apply the general AC power flow model, the small signal model is determined by the steady-state variables, and further determined by the DG droop set point $\bm{z}=[\bm{P}^{*T}_G, \bm{Q}^{*T}_G, \bm{V}^{*T}]^T \in \mbR^{3g}$ and $\bm{\epsilon}_{R}=\{\epsilon_{Ri}\} \in \mbR^{d}$. The small signal model is given as:
\begin{eqnarray} \label{SDM}
\left[ \begin{matrix}
    \Delta \dot{\bm{x}}\\\textbf{0}
\end{matrix} \right]
=\left[ \begin{matrix}
	\bm{A}&		\bm{B}\\
	\bm{C}&		\bm{D}\\
\end{matrix} \right]
\left[ \begin{matrix}
	\Delta  \bm{x}\\
	\Delta  \bm{y}\\
\end{matrix} \right]
\end{eqnarray}
where the sub-matrices $\bm{A}\in\mbR^{(3g-1)\times{(3g-1)}}$ and $\bm{B}\in\mbR^{(3g-1)\times{2d}}$ represent differential equations \eqref{angledynamics}-\eqref{Gvoldynamics} with respect to state variables and algebraic variables, respectively; and $\bm{C}\in\mbR^{2d\times{(3g-1)}}$ and $\bm{D}\in\mbR^{2d\times{2d}}$ represent algebraic equations \eqref{loadPdynamics}-\eqref{loadQdynamics} with respect to state variables and algebraic variables, respectively. By the system dynamics, the sub-matrices $\bm{A}$, $\bm{B}$, $\bm{C}$, $\bm{D}$ in \eqref{SDM} are formulated in
\begin{eqnarray}\label{dynamicalmatrix}
\left[
	\begin{array}{c:c}
	\bm{A}& \bm{B} \\ \hdashline
	\bm{C}& \bm{D}
	\end{array}
	\right]=
\left[
    \begin{array}{ccc : cc}
   \textbf{0} &  \omega_b\bm{T}_G  & \textbf{0} & \textbf{0} & \textbf{0}  \\
   \frac{\partial f_G}{\partial \bm{\alpha}_r}  &  -\bm{F_p}\bm{I_{g}}  &  \frac{\partial f_G}{\partial \bm{V}_G} & \frac{\partial f_G}{\partial \bm{\alpha}_L} & \frac{\partial f_G}{\partial \bm{V}_L} \\
   \frac{\partial h_G}{\partial \bm{\alpha}_r}  & \textbf{0} & \frac{\partial h_G}{\partial \bm{V}_G} & \frac{\partial h_G}{\partial \bm{\alpha}_L} & \frac{\partial h_G}{\partial \bm{V}_L}\\ \hdashline
   \frac{\partial f_L}{\partial \bm{\alpha}_r} & \textbf{0} & \frac{\partial f_L}{\partial \bm{V}_G} &  \frac{\partial f_L}{\partial \bm{\alpha}_L} & \frac{\partial f_L}{\partial \bm{V}_L}\\
   \frac{\partial h_L}{\partial \bm{\alpha}_r} & \textbf{0} & \frac{\partial h_L}{\partial \bm{V}_G} & \frac{\partial h_L}{\partial \bm{\alpha}_L} & \frac{\partial h_L}{\partial \bm{V}_L}\\
   \end{array}
   \right]
\end{eqnarray}
where ${\bm{T}_G}= \left[ { - {\bm{1}_{g-1}},{\bm{I}_{g-1}}} \right]$; the power flow Jacobian entries with partial differential operators can be calculated from the power flow model at the equilibrium point $(\bm{x}^0,\bm{y}^0)$, which determined by $(\bm{z},\bm{\epsilon}_R)$. Thus, the sub-matrices $\bm{A}$, $\bm{B}$, $\bm{C}$, $\bm{D}$ are also dependent on $(\bm{z},\bm{\epsilon}_R)$. Due to the space limit, we will omit the explicit expressions of the aforementioned Jacobian matrices with partial differential operators, such as $\frac{\partial f_G}{\partial \bm{\alpha}_r}$ in this paper, readers can refer to \cite{b23} to obtain the detailed expression of power flow Jacobian entries.

It is common that the sub-matrix $\bm{D}$ in \eqref{SDM} is nonsingular \cite{b24}. Then the dynamic Jacobian matrix $\bm{J}$ is given by eliminating $\Delta \bm{y}$
\begin{equation}
{\Delta \dot{\bm{x}}} = (\bm{A} - \bm{B}{\bm{D}^{ - 1}}\bm{C})\bm{x}={\bm{J}}\Delta \bm{x}.
\end{equation}

Referring to the definition of the stability index in our previous work \cite{b25}, we describe the stability index $\eta$ in term of the following semi-definite programming (SDP) problem, which is friendly to form an analytical stability sensitivity. What should be noted is that $\eta$ depends on the equilibrium point, and further relies on $(\bm{z},\epsilon_R)$.
\begin{subequations}\label{stabilityindexdef}
\begin{align}
   \min &\quad {\eta}\\
   \textup{s.t.} &\quad - {\bm{J}^T}{\bm{\Phi}} - {\bm{\Phi}}{\bm{J}}+ {\eta}\bm{I} \succeq 0\\
   &\quad {\bm{\Phi}} - \epsilon \bm{I}\succeq 0\label{existence}\\
   &\quad  -{\bm{\Phi}} + \bm{I}\succeq 0\label{uni}
   \vspace{-10pt}
\end{align}
\end{subequations}
where $\bm{\Phi}$ is a symmetric semi-definite matrix. Based on the above definition of the stability index, the absolute value of stability index $|\eta|$ is the lower bound of system oscillation convergence rate. The
curve $e^{\eta t}$ is the exponential envelope of the system oscillation under small disturbances. In the conventional view without considering the uncertainties, the stability index is restricted by a negative upper band,
\begin{equation}
\label{constabilitycon}
\eta \leq \bar{\eta} < 0.
\end{equation}
\vspace{-20pt}
\section{CC-OPF Model Accounting Small Signal Stability}
Due to the uncertainties brought by RES forecast errors, conventional deterministic optimization frameworks become inadequate for handling the risks of voltage violation, generation violation as well as the stability issue. Hence, CCs are adopted to ensure the constraints under corresponding confidence levels. In the following, we will present a bi-level CC-OPF in one timeslot (whose duration is 15 minutes) which incorporates the stability constraint,
\begin{subequations}
\begin{align}
 \mathop { \min }\limits & \ \ \mbE_F[\sum\nolimits_{i \in \mV_G}a_{2i}P_{Gi}^2+ a_{1i}P_{Gi}+a_{0i}]\label{obj}\\
 \textup{s.t.} & \quad \eqref{powerinj}\\
 & \quad \mbP_F\{P_{Gi} \leq \bar{P}_{Gi}\}\geq 1-\beta_G, i \in \mV_G\label{chanceconDGPup}\\
 & \quad \mbP_F\{P_{Gi} \geq \uline{P}_{Gi}\}\geq 1-\beta_G, i \in \mV_G\label{chanceconDGPlo}\\
 & \quad \mbP_F\{Q_{Gi} \leq \bar{Q}_{Gi}\}\geq 1-\beta_G, i \in \mV_G\label{chanceconDGQup}\\
 & \quad \mbP_F\{Q_{Gi} \geq \uline{Q}_{Gi}\}\geq 1-\beta_G, i \in \mV_G\label{chanceconDGQlo}\\
 &\quad \mbP_F\{V_i \leq \bar{V}_i\}\geq 1-\beta_V, i \in \mV_G \cup \mV_L\label{chanceconVup}\\
 &\quad \mbP_F\{V_i \geq \uline{V}_i\}\geq 1-\beta_V, i \in \mV_G \cup \mV_L\label{chanceconVlo}\\
 & \quad \mbP_F\{\eta(\bm{z},\bm{\epsilon_R}) \leq \bar{\eta}\}\geq 1-\beta_{\eta}\label{chanceconstabillity}\\
 & \quad \eta(\bm{z},\bm{\epsilon_R}) = \min \phi\\
 & \quad \quad \quad \quad \quad \quad \textup{s.t.}  - {\bm{J}^T}{\bm{\Phi}} - {\bm{\Phi}}{\bm{J}}+ {\phi}\bm{I} \succeq 0\\
 &  \quad \quad \quad \quad \quad \quad \quad {\bm{\Phi}} - \epsilon \bm{I}\succeq 0\label{existence}\\
 &  \quad \quad \quad \quad \quad \quad  \quad  -{\bm{\Phi}} + \bm{I}\succeq 0\label{uni}
\end{align}
\end{subequations}
where $a _{0i}$, $a _{1i}$ and $a _{2i}$ in \eqref{obj} are fuel cost coefficients; $F$ denotes the PDF of the forecast error $\bm{\epsilon}_R$, which will be estimated in next section; chance constraints \eqref{chanceconDGPup}-\eqref{chanceconVlo} limit the probability of the DG outputs and voltage magnitude within the upper and lower bounds greater than or equal to the confidence level $1-\beta_G$ or $1-\beta_V$; while the chance constraint for the stability index \eqref{chanceconstabillity} also ensures the probability of the system small signal stability greater than the prescribed value $1-\beta_{\eta}$, which brings more reliable solutions with sufficient stability against RES forecast errors. The decision variables in the proposed model include the stability index $\eta$, DG droop set points $\bm{z}$ and other operating variables $(\bm{P}_G, \bm{Q}_G, \bm{x}, \bm{y})$. According to the physics in power systems, the decision variables should be separated into two categories, independent variables $\bm{z}$ and dependent variables $\eta$, $\bm{P}_G, \bm{Q}_G, \bm{x}, \bm{y}$. Dependent variables can be seen as implicit functions w.r.t. independent variables and RES forecast error $\bm{\epsilon}_R$ via power flow equation. Note that the CCs in (9) are with respect to dependent variables, in the next subsection we will reformulate them in terms of independent variables.

The proposed CC-SC-OPF cannot be directly resolved by solvers without reformulations. Traditional linear power flow based CC-OPF without considering the stability CC can be easily transformed into a SOCP under assumed Gaussian uncertainties. Nevertheless, the proposed model adopts the ACPF and integrates the stability CC. Moreover, the RES uncertainties are relaxed to follow non-Gaussian distributions. Because of the nonlinearity of ACPF and stability index and non-Gaussian type of uncertainty, the proposed model is much more complicated. Then, the general idea of the solution method is introduced as follows. Firstly, we apply the Gaussian mixture model (GMM) to fit the PDF of non-Gaussian uncertainties. Secondly, we incorporate the stability CC into CC-OPF by linearizing the implicit function between $\eta$ and $(\bm{z},\bm{\epsilon_R})$ through adopting the first-order Taylor expansion with analytical sensitivity. The technical details will be clarified in the next section.


\section{GMM-based Uncertainty Modeling and Chance Constraint Reformulations}
\subsection{Non-Gaussian Uncertainty Modeling based on GMM}
To reformulate the aforementioned CCs, we should determine the joint PDF, CDF and even inverse CDF of the uncertainties. But, the RES uncertainties are represented by forecast error $\bm{\epsilon}_R$, whose entries are non-Gaussian correlated random variables. How to accurately estimate the joint distribution of correlated non-Gaussian variables is hard to deal with. Monte-Carlo simulation and Gaussian estimation, which are widely applied to resolve the challenge, are time consuming or inaccurate. On the contrary, Gaussian mixture model (GMM) has the benefit on fitting the distribution efficiently through a linear combination of multi-dimensional Gaussian distribution functions $N(\bm{x}; \bm{\mu}_m, \bm{\sigma}_m)$. Moreover, GMM inherits good properties from Gaussian distributions, such as linear invariance, which brings convenience to transform CCs \cite{b26}. Therefore, we adopt GMM to estimate the distribution of non-Gaussian forecast error $\bm{\epsilon}_R$,
\begin{subequations}
\begin{align}
\textup{PDF}_{\mE}(\bm{\epsilon}_R) &= \sum_{m=1}^M \omega_m N(\bm{\epsilon}_R; \bm{\mu}_m, \bm{\sigma}_m)\\
\textup{CDF}_{\mE}(\bm{\epsilon}_R) &= \sum_{m=1}^M \omega_m [\int_{\bm{v} \leq {\epsilon}_R}N(\bm{v}; \bm{\mu}_m, \bm{\sigma}_m)d\bm{v}]\\
\textup{where} \quad N(\bm{\epsilon}_R; &\bm{\mu}_m, \bm{\sigma}_m) = \frac{e^{-\frac{1}{2}(\bm{\epsilon_R}-\bm{\mu_m})^T\bm{\Sigma}_m^{-1}(\bm{\epsilon_R}-\bm{\mu_m})}}{(2\pi)^{\frac{d}{2}}\textup{det}(\bm{\Sigma}_m)^{\frac{1}{2}}} \nonumber \\
\sum_{m=1}^M\omega_m& = 1 \nonumber
\end{align}
\end{subequations}
where $M$ is the number of Gaussian components, to avoid the overfitting or underfitting issues, after several validation tests, we select $M=10$; the coefficient $\omega_m$, expectance vector $\bm{\mu}_m$ and covariance matrix $\bm{\sigma}_m$ are adjustment parameters. With RES historical data in hand, we estimate these parameters by commonly used maximum likelihood estimation (MLE). Actually, applying MLE in the GMM parameter fitting problem is quite mature and practical, which has been integrated in some commercial software, such as: \textbf{fitgmdist} function in MATLAB.

\subsection{Reformulations of CCs on Stability, Voltage, and DG Generation}
\subsubsection{Linear Approximation of Stability Index}
In addition to handling security constraints as in the existing CC-OPF models, this paper aims to ensure the small-signal stability under uncertain RES outputs. The biggest challenge for transforming the stability CC is its high nonlinearity and implicit expressions w.r.t. variables $(\bm{z},\bm{\epsilon}_R)$. Note that the proposed CC-SC-OPF is mainly used to determine the dispatch scheme for the upcoming timeslot, e.g., the 15 min-ahead dispatch. The RES forecast error under this timescale (especially for the wind power) is less than 5 \% according to some advanced prediction approaches, such as, sparse vector auto-regression in \cite{b27}. Hence, the linearization approach is proper to address this challenge. The most straightforward approach to establish the linearized model is the numerical-perturbation based approach, which directly calculates the sensitivity of the probability of $\eta$ within the constraint to $(\bm{z},\bm{\epsilon}_R)$.
\begin{equation}
\begin{split}
\mP &= \mP(\bm{\xi}_0)+ \frac{\partial \mP}{\partial \bm{\xi}} \cdot (\bm{\xi}-\bm{\xi}_0)\\
\frac{\partial \mP}{\partial \bm{\xi}}&\simeq \frac{\mP(\bm{\xi}_0+e_{\eta})-\mP(\bm{\xi}_0)}{e_{\eta}}
\end{split}
\label{nup}
\end{equation}
where $\mP=\mbP\{\eta(z,\epsilon_R) \leq \bar{\eta}\}$, $\bm{\xi}=(\bm{z},\bm{\epsilon}_R)$, $\bm{\xi}_0$ denotes the decision variable $(\bm{z},\bm{\epsilon}_R)$ at the equilibrium point, $e_{\eta}$ is the amount of perturbation. This approach is highly time consuming because it needs one more round of Monte-Carlo simulations to only obtain the sensitivity w.r.t. a single variable. This makes the numerical perturbation impractical to be adopted in the CC-SC-OPF which needs to output the solution within a short timeslot.

To overcome these obstacles, we propose to use analytical sensitivity to establish the linearly approximated relation between the stability index $\eta$ and $\bm{\xi}$, which avoids the time-consuming Monte Carlo simulations in numerical perturbation. Note that $\eta$ is the optimal objective value of the lower level problem of (9), which is an implicit function of $\xi$. Thus, we have the first order Taylor expansion around the equilibrium point $\eta_0$ under the RES forecast output $\bm{\bar{P}_{R}}$ and corresponding set point in DG droop controllers $\bm{P^{*}_{G}}$, $\bm{Q^{*}_{G}}$, $\bm{V^*_{G}}$ as well as the forecast error $\bm{\epsilon}_R$,
\begin{equation}\label{etasensitivity}
\begin{split}
\eta \simeq & \ \eta_0 + \frac{\partial \eta}{\partial \bm{P^{*}_{G}}} \Delta \bm{P^{*}_{G}}+\frac{\partial \eta}{\partial \bm{Q^{*}_{G}}} \Delta \bm{Q^{*}_{G}}+\frac{\partial \eta}{\partial \bm{V^{*}_{G}}} \Delta \bm{V^{*}_{G}}\\
&+ \left (\frac{\partial \eta}{\partial \bm{P_{R}}}+\lambda \frac{\partial \eta}{\partial \bm{Q_{R}}}\right) \bm{\epsilon}_R
\end{split}
\end{equation}
where the partial derivative terms are stability sensitivities. In the following we derive the formula for these sensitivity terms. Take the sensitivity of $\eta$ to DG set active power $\frac{\partial \eta}{\partial \bm{P^{*}_{G}}}$ as an example and the formula for other sensitivity terms can be obtained in a similar way.

By applying the chain rule, the sensitivity of stability index to $\bm{P^{*}_{G}}$ can be explicitly formulated,
\vspace{-5pt}
\begin{equation}\label{anasensitivity}
\begin{split}
&\frac{\partial \eta}{\partial \bm{P^{*}_{G}}} =\sum_{m,n}\frac{\partial \eta}{\partial {J_{mn}}}\cdot \frac{\partial {J_{mn}}}{\partial \bm{P^{*}_{G}}}=\\
&\!\sum_{m,n}\!\frac{\partial \eta}{\partial {J_{mn}}}\!\left(\!\sum_{i \in \mV}\!\frac{\partial {J_{mn}}}{\partial \alpha_i}\!\frac{\partial \alpha_i}{\partial \bm{P^{*}_{G}}}\!\!+\!\!\frac{\partial {J_{mn}}}{\partial V_i}\!\frac{\partial V_i}{\partial \bm{P^{*}_{G}}}\!\!+\!\! \frac{\partial {J_{mn}}}{\partial \omega_i}\! \frac{\partial \omega_i}{\partial \bm{P^{*}_{G}}}\!\!\right)
\end{split}
\end{equation}
where $J_{mn}$ denotes any entry in the Jacobian matrix, which is determined directly by operational variables $(\bm{x},\bm{y})$. The analytical formula for the sensitivity of the stability index to the Jacobian entry $\frac{\partial \eta}{\partial J_{mn}}$ has been established in our previous work \cite{b25} by leveraging the strong duality held by the convex lower level problem of (9). The details are omitted here due to page limit. We only need to focus on the partial derivative in the inner bracket of right hand side (RHS) in \eqref{anasensitivity}.

Firstly, we consider the partial derivatives in dynamical states $\frac{\partial {J_{mn}}}{\partial \alpha_i}$, $\frac{\partial {J_{mn}}}{\partial V_i}$, $\frac{\partial {J_{mn}}}{\partial \omega_i}$. According to the system small-signal stability analysis, the dynamic Jacobian matrix can be formulated by four sub-matrices in \eqref{dynamicalmatrix}. Each entry of these sub-matrices has been explicitly expressed. Hence partial derivatives $\frac{\partial {J_{mn}}}{\partial \alpha_i}$, $\frac{\partial {J_{mn}}}{\partial V_i}$, $\frac{\partial {J_{mn}}}{\partial \omega_i}$ can be easily formulated without much additional computations.

Secondly, we need to construct the analytical expression of partial derivatives in static states $\frac{\partial \alpha_i}{\partial \bm{P^{*}_{G}}},\frac{\partial V_i}{\partial \bm{P^{*}_{G}}},\frac{\partial \omega_i}{\partial \bm{P^{*}_{G}}}$. From the system-wide DAE \eqref{dynamicmodel}, the explicit functions of the state variables $(\bm{x},\bm{y})$ to the DG droop set point $(\bm{P}_{G}^*,\bm{Q}_{G}^*)$ can be directly derived through the sensitivity matrix $\bm{S}$. Let the left hand side (LHS) of \eqref{dynamicmodel} be zero to obtain equilibrium equations and derive corresponding matrix form of incremental formulations:
\begin{eqnarray} \label{SDM}
\setlength{\arraycolsep}{1pt}
\hspace{-8mm}
\left[ \begin{matrix}
    \Delta \bm{P}_{G}^*\\\Delta \bm{Q}_{G}^*
\end{matrix} \right]
\!\!=\!\left[ \begin{matrix}
	\frac{\partial \bm{P}^\textup{inj}}{\partial \bm{\alpha}_r}&\frac{\partial \bm{P}^\textup{inj}}{\partial \bm{\alpha}_L}&\frac{\partial \bm{P}^\textup{inj}}{\partial \bm{V}}& \bm{K}_{p}^{-1}\\
	\frac{\partial \bm{Q}^\textup{inj}}{\partial \bm{\alpha}_r}&\frac{\partial \bm{Q}^\textup{inj}}{\partial \bm{\alpha}_L}&\frac{\partial \bm{Q}^\textup{inj}}{\partial \bm{V}}+\bm{K}_{q}^{-1}&		\bm{0}\\
\end{matrix} \right]
\left[ \begin{matrix}
	\Delta  \bm{\alpha}_r\\
    \Delta  \bm{\alpha}_L\\
	\Delta  \bm{V}\\
    \Delta  \omega
\end{matrix} \right]\!\!=\!
\bm{S} \left[ \begin{matrix}
	\Delta  \bm{\alpha}_r\\
    \Delta  \bm{\alpha}_L\\
	\Delta  \bm{V}\\
    \Delta  \omega
\end{matrix} \right]
\end{eqnarray}
where $\bm{P}_{G}^*=\{P_{Gi}^*\} \in \mbR^n$, if $i \in \mV_L$, set $\bm{P}_{Gi}^*=0$, and similar definition to $\bm{Q}_{G}^*$; $\bm{K}_{p}^{-1} =\{K_{pi}^{-1}\} \in \mbR^{n}$, $\bm{K}_{q}^{-1} \in \mbR^{n \times n}$ is diagonal matrix that collects $K_{qi}^{-1}$ as its main diagonals; the state vector $[\bm{\alpha}_r^T, \bm{\alpha}_L^T,\bm{V}^T, \bm{\omega}]^T \in \mbR^{2n}$. Hence, the sensitivity matrix $\bm{S} \in \mbR^{2n \times 2n}$, and in general cases the matrix $\bm{S}$ is invertible. Accordingly, the analytical expression of partial derivatives $\frac{\partial \alpha_i}{\partial \bm{P^{*}_{G}}},\frac{\partial V_i}{\partial \bm{P^{*}_{G}}},\frac{\partial \omega_i}{\partial \bm{P^{*}_{G}}}$ are corresponding entries in the inverse DAE matrix $\bm{S}^{-1}$.

By the aforementioned derivative constructions, we are ready to present the partial derivative of the stability index $\eta$ to $\bm{P}^*_G$ in \eqref{anasensitivity}, and similar sensitivity reconstructions apply to other partial derivatives of the stability index to its decision variables, such as $\frac{\partial \eta}{\partial \bm{Q^{*}_{G}}}, \frac{\partial \eta}{\partial \bm{V^{*}_{G}}}$ and $\frac{\partial \eta}{\partial \bm{P_{R}}}, \frac{\partial \eta}{\partial \bm{Q_{R}}}$, where
\begin{subequations}
\begin{align}
\frac{\partial \eta}{\partial \bm{Q^{*}_{G}}} =\sum_{m,n}\frac{\partial \eta}{\partial {J_{mn}}}\cdot \frac{\partial {J_{mn}}}{\partial \bm{Q^{*}_{G}}}\\
\frac{\partial \eta}{\partial \bm{V^{*}_{G}}} =\sum_{m,n}\frac{\partial \eta}{\partial {J_{mn}}}\cdot \frac{\partial {J_{mn}}}{\partial \bm{V^{*}_{G}}}\\
\frac{\partial \eta}{\partial \bm{P_{R}}} =\sum_{m,n}\frac{\partial \eta}{\partial {J_{mn}}}\cdot \frac{\partial {J_{mn}}}{\partial \bm{P_{R}}}\\
\frac{\partial \eta}{\partial \bm{Q_{R}}} =\sum_{m,n}\frac{\partial \eta}{\partial {J_{mn}}}\cdot \frac{\partial {J_{mn}}}{\partial \bm{Q_{R}}}.
\end{align}
\end{subequations}

The above analytical formulae for the sensitivity terms provide a more accurate and efficient way than numerical perturbation for evaluating the sensitivities in the construction of linear approximation of stability CC. The effectiveness of the proposed linear approximation will be verified in the case study section.

\subsubsection{Reconstruction of Stability CC}

Owing to the linear approximation of the stability index, the stability CC \eqref{chanceconstabillity} can be rewritten as:
\begin{equation}
\hspace{-2mm}
\mbP
\left\{\begin{array}{ll}
             \eta_0 +\frac{\partial \eta}{\partial \bm{P^{*}_{G}}} \Delta \bm{P^{*}_{G}}+\frac{\partial \eta}{\partial \bm{Q^{*}_{G}}} \Delta \bm{Q^{*}_{G}}+\frac{\partial \eta}{\partial \bm{V^{*}_{G}}} \Delta \bm{V^{*}_{G}}\\
             + (\frac{\partial \eta}{\partial \bm{P_{R}}}+\lambda \frac{\partial \eta}{\partial \bm{Q_{R}}}) \bm{\epsilon}_R \leq \bar{\eta}
             \end{array}
\right\}\geq 1-\beta_{\eta}
\label{CCSC}
\end{equation}
where the sensitivity values are given by the formulae in last subsection. Since the PDF and CDF of the RES forecast error are already given by GMM approach, we convert \eqref{CCSC} in terms of $\bm{\epsilon}_R$.
For the notation simplicity, we define a new random variable $\mU_{R}$ by $\mU_{R} = (\frac{\partial \eta}{\partial \bm{P_{R}}}+\lambda \frac{\partial \eta}{\partial \bm{Q_{R}}}) \bm{\epsilon}_R=\bm{\alpha}^T\bm{\epsilon}_R$.
Based on the definition of CDF,
we obtain the following linear constraint as an approximation of the stability CC around a certain solution, which helps to enforce the stability CC in a more tractable way,
\begin{equation}
\! \bar{\eta} -\bigg(\eta + \frac{\partial \eta}{\partial \bm{P^{*}_{G}}} \! \Delta \bm{P^{*}_{G}}+\frac{\partial \eta}{\partial \bm{Q^{*}_{G}}} \Delta \bm{Q^{*}_{G}}+\frac{\partial \eta}{\partial \bm{V^{*}_{G}}} \Delta \bm{V^{*}_{G}}\bigg) \! \geq \! \textup{CDF}^{-1}_{\mU_{R}}(1-\beta_{\eta})
\label{reformulatedCCSC}
\end{equation}
where $\textup{CDF}^{-1}_{\mU_{R}}(\cdot)$ is inverse CDF of the new random variable $\mU_{R}$, and $\textup{CDF}^{-1}_{\mU_{R}}(1-\beta_{\eta})$ is a constant number, the calculation method of which will be detailed in the following. It is worth noting that \eqref{reformulatedCCSC} will work as Benders cut in the solution methodology, which will enforce the stability index to satisfy the CC. The solution method will be given in the next section.

Because of the perfect linear invariance property of the GMM estimation approach, the PDF and CDF of random variable $\mU_{R}$ can be deduced as:
\begin{subequations}\label{PDFCDFeta}
\begin{align}
&\textup{PDF}_{\mU_R}(\bm{u}_R) = \sum_{m=1}^M \omega_m N(\bm{u}_R; \bm{\alpha}^T\bm{\mu}_m, \bm{\alpha}^T\bm{\sigma}_m\bm{\alpha})\\
&\textup{CDF}_{\mU_R}\!(\!\bm{u}_R\!) \!= \!\!\!\sum_{m=1}^M \!\omega_m \![\int_{\bm{w}\leq \bm{u}_R}\!\!\!\!\!\!N(\bm{w};\! \bm{\alpha}^T\bm{\mu}_m, \bm{\alpha}^T\!\bm{\sigma}_m\bm{\alpha})d\bm{w}]
\end{align}
\end{subequations}
where the fitting parameters $\omega_m, \bm{\mu}_m,\bm{\sigma}_m$ are estimated according to the RES historical data. According to the explicit expression of the CDF, $\textup{CDF}^{-1}_{\mU_{R}}(\cdot)$ can be conveniently calculated in MATLAB. During the entire optimization process, we only need to estimate the density function of the RES uncertainty once. Consequently, the computational complexity will dramatically reduce compared to the pure numerical perturbation approach based on Monte-Carlo simulations as shown in \eqref{nup}.
\subsubsection{Reconstructions of Security CC}
Similar to the reconstruction of the stability CC, the expression of bus voltages and DG generations w.r.t. $(\bm{z},\bm{\epsilon}_R)$ also should be linearized. Fortunately, it is beneficial that the linearization degree of the voltage/DG generation to the DG droop set point/ RES outputs are sufficiently higher than the stability index. In other words, the reformulations of bus voltage/DG generations CCs by using the partial derivative operators will be more accurate than the stability index. The reformulated formulae for \eqref{chanceconVlo}-\eqref{chanceconVlo} are presented as:
\begin{subequations}
\begin{align}
\hspace{-20mm}
&\!\!\!\!\!\bar{P}_{Gi} \!\!- \!\!\bigg({P}_{Gi}\! \!+\!\!\frac{\partial {P}_{Gi}}{\partial \bm{P^{*}_{G}}} \!\!\Delta \bm{P^{*}_{G}}\!\!+\!\!\frac{\partial {P}_{Gi}}{\partial \bm{Q^{*}_{G}}} \!\!\Delta \bm{Q^{*}_{G}}\!\!+\!\!\frac{\partial {P}_{Gi}}{\partial \bm{V^{*}_{G}}} \!\Delta \bm{V^{*}_{G}}\bigg) \!\geq \!\textup{CDF}^{-1}_{\mU_{Pi}}\!(1\!-\!\!\beta_{G}\!)\\
&\!\!\!\!\! \!\bigg({P}_{Gi} \!\!+\!\!\frac{\partial {P}_{Gi}}{\partial \bm{P^{*}_{G}}}\! \!\Delta \bm{P^{*}_{G}}\!\!+\!\frac{\partial {P}_{Gi}}{\partial \bm{Q^{*}_{G}}} \!\!\Delta \bm{Q^{*}_{G}}\!\!+\!\!\frac{\partial {P}_{Gi}}{\partial \bm{V^{*}_{G}}} \!\!\Delta \bm{V^{*}_{G}}\bigg)\!\!-\!\!\uline{P}_{Gi} \!\geq\!\textup{CDF}^{-1}_{\mU_{Pi}}\!(1\!-\!\!\beta_{G}\!)\\
&\!\!\!\!\!\bar{Q}_{Gi} \!\!- \!\!\bigg({Q}_{Gi}\! \!+\!\!\frac{\partial {Q}_{Gi}}{\partial \bm{P^{*}_{G}}} \!\!\Delta \bm{P^{*}_{G}}\!\!+\!\!\frac{\partial {Q}_{Gi}}{\partial \bm{Q^{*}_{G}}} \!\!\Delta \bm{Q^{*}_{G}}\!\!+\!\!\frac{\partial {Q}_{Gi}}{\partial \bm{V^{*}_{G}}} \!\Delta \bm{V^{*}_{G}}\bigg) \!\geq \!\textup{CDF}^{-1}_{\mU_{Qi}}\!(1\!-\!\!\beta_{G}\!)\\
&\!\!\!\!\! \!\bigg({Q}_{Gi} \!\!+\!\!\frac{\partial {Q}_{Gi}}{\partial \bm{P^{*}_{G}}}\! \!\Delta \bm{P^{*}_{G}}\!\!+\!\frac{\partial {Q}_{Gi}}{\partial \bm{Q^{*}_{G}}} \!\!\Delta \bm{Q^{*}_{G}}\!\!+\!\!\frac{\partial {Q}_{Gi}}{\partial \bm{V^{*}_{G}}} \!\!\Delta \bm{V^{*}_{G}}\bigg)\!\!-\!\!\uline{Q}_{Gi} \!\geq\!\textup{CDF}^{-1}_{\mU_{Qi}}\!(1\!-\!\!\beta_{G}\!)\\
&\!\!\!\!\!\bar{V}_i \!\!- \!\!\bigg(V_{i}\! \!+\!\!\frac{\partial V_i}{\partial \bm{P^{*}_{G}}} \!\!\Delta \bm{P^{*}_{G}}\!\!+\!\!\frac{\partial V_i}{\partial \bm{Q^{*}_{G}}} \!\!\Delta \bm{Q^{*}_{G}}\!\!+\!\!\frac{\partial V_i}{\partial \bm{V^{*}_{G}}} \!\Delta \bm{V^{*}_{G}}\bigg) \!\geq \!\textup{CDF}^{-1}_{\mU_{V_i}}\!(1\!-\!\!\beta_{V}\!)\\
&\!\!\!\!\! \!\bigg(V_{i} \!\!+\!\!\frac{\partial V_i}{\partial \bm{P^{*}_{G}}}\! \!\Delta \bm{P^{*}_{G}}\!\!+\!\frac{\partial V_i}{\partial \bm{Q^{*}_{G}}} \!\!\Delta \bm{Q^{*}_{G}}\!\!+\!\!\frac{\partial V_i}{\partial \bm{V^{*}_{G}}} \!\!\Delta \bm{V^{*}_{G}}\bigg)\!\!-\!\!\uline{V}_i \!\geq\!\textup{CDF}^{-1}_{\mU_{V_i}}\!(1\!-\!\!\beta_{V}\!).
\end{align}
\end{subequations}
where new random variable $\mU_{P},\mU_{Q},\mU_{V}$ are defined by $\mU_{Pi} = (\frac{\partial P_{Gi}}{\partial \bm{P_{R}}}+\lambda \frac{\partial P_{Gi}}{\partial \bm{Q_{R}}}) \bm{\epsilon}_R$, $\mU_{Qi} = (\frac{\partial Q_{Gi}}{\partial \bm{P_{R}}}+\lambda \frac{\partial Q_{Gi}}{\partial \bm{Q_{R}}}) \bm{\epsilon}_R$, $\mU_{Vi} = (\frac{\partial V_i}{\partial \bm{P_{R}}}+\lambda \frac{\partial V_i}{\partial \bm{Q_{R}}}) \bm{\epsilon}_R$. Because of the linear invariance property of GMM, similar to the construction of the CDF of $\mU_R$ \eqref{PDFCDFeta}, the inverse CDF of $\mU_{Pi}, \mU_{Qi}, \mU_{Vi}$ can be deduced.

\section{Solution Methodology}
\subsection{Benders Decomposition based Solution Approach}
With the aforementioned reformulations, CCs in the proposed OPF are transformed into a bi-level problem with deterministic linear constraints. In particular, the stability CC associated with the lower level problem is reformulated in (17). Except the stability CC, the upper level problem is a quadratic program (QP) and lower level problem is a SDP problem, which is hard to be incorporated in the upper level through Karush-Kuhn-Tucker (KKT) conditions. Hence, the upper level and lower level problems should be solved separately. The Benders decomposition (BD) incorporates the feature of the proposed OPF model and the stability CC can be comprised through Benders cut (17).

By applying the BD based approach, the bi-level problem is decomposed into a master problem and subproblems, where the master problem contains the upper level excluding the stability constraint; and the subproblems consist of the lower level problem and the stability CC. Take the $k^{\textup{th}}$ iteration as an instance to illustrate the solution method.

\textit{\textbf{Step 1}} Solve the master problem and obtain the optimal solution $\bm{d}_\textup{opt}^{(k)}$, where $\bm{d}=[\bm{P_G}^T, \bm{Q_G}^T, \bm{x}, \bm{y}, \bm{z}]^T$.

\textit{\textbf{Step 2}} With the optimum $\bm{d}_\textup{opt}^{(k)}$ as known parameter, solve the lower level problem and obtain the stability index ${\eta}^{(k)}$.

\textit{\textbf{Step 3}} Apply the reformulated stability CC (20) with the optimized independent variable $\bm{z}_{\textup{opt}}^{(k)}$
\begin{equation}
\bar{\eta}-\eta^{(k)} \geq \textup{CDF}^{-1}_{\mU_{R}}(1-\beta_{\eta}).
\label{reformulatedCCSCfixed}
\end{equation}
to check if the obtained ${\eta}^{(k)}$ is admissible. In case of the stability CC violation, generate the Benders cut \eqref{Benderscut},
\begin{equation}\label{Benderscut}
{\eta}^{(k)} +\frac{\partial \eta}{\partial \bm{z}}\bigg|_{\bm{d}_\textup{opt}^{(k)}} (\bm{z}^{(k+1)}-\bm{z}_\textup{opt}^{(k)}) \leq \bar{\eta}-\textup{CDF}^{-1}_{\mU_{R}}(1-\beta_{\eta}).
\end{equation}

\textit{\textbf{Step 4}} On the basis of the Benders cut, construct additional linear constraints to update the master problem. The iteration stops when all constraints are satisfied.

\begin{figure}[htbp]
\centerline{\includegraphics[width=3in]{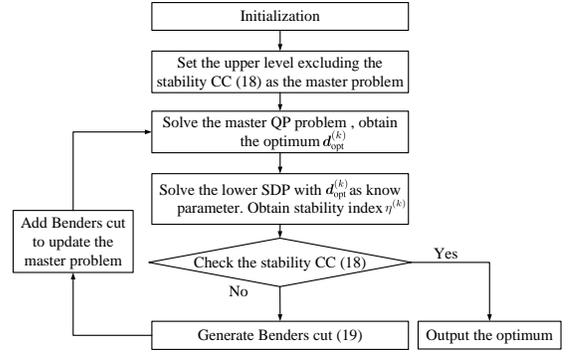}}
\caption{The flowchart of Benders based solution methodology}
\label{bendersflowchart}
\end{figure}

The above manipulations is summarized in Fig.~\ref{bendersflowchart}.

\subsection{Supplementary Corrective Countermeasure}
Because the proposed optimization model involves the probabilistic behavior described by AC power flow, without the linearization, CCs cannot be reformulated into linear deterministic forms. Although the linearization is easy to implement with high efficiency, it inevitably brings small errors in the optimization process and may bring a few exceptions to the solution. These exceptions mean even though the reformulated linear constraints are satisfied, the original CCs are violated. To cope with the defect, this subsection provides a supplementary corrective countermeasure.

Take the stability CC as an example to illustrate the supplementary countermeasure. The linear approximation ofnstability CC in the last iteration is satisfied, we suppose that the last iteration is $l^\textup{th}$ iteration,
\begin{equation}
\begin{split}
\bar{\eta} -\bigg(&\eta^{(l)} + \frac{\partial \eta}{\partial \bm{P^{*}_{G}}}\bigg|_{\bm{d}_\textup{opt}^{(l)}} \! \Delta \bm{P^{*}_{G}}+\frac{\partial \eta}{\partial \bm{Q^{*}_{G}}}\bigg|_{\bm{d}_\textup{opt}^{(l)}} \Delta \bm{Q^{*}_{G}}
\\&+\frac{\partial \eta}{\partial \bm{V^{*}_{G}}}\bigg|_{\bm{d}_\textup{opt}^{(l)}} \Delta \bm{V^{*}_{G}}\bigg)
\geq \textup{CDF}^{-1}_{\mU_{R}}(1-\beta_{\eta})
\end{split}
\end{equation}
while the original CC $\mbP_F\{\eta^{(l)}-\bar{\eta}\} \geq 1-\beta_{\eta}$ with the optimum is violated. The supplementary corrective countermeasure includes following steps,

\textit{\textbf{Step 1}} Calculate the absolute values of partial derivative terms $|\frac{\partial \eta}{\partial \bm{P^{*}_{G}}}|$, $|\frac{\partial \eta}{\partial \bm{Q^{*}_{G}}}|$, $|\frac{\partial \eta}{\partial \bm{V^{*}_{G}}}|$ at the optimum $\bm{d}^{(l)}_\textup{opt}$.

\textit{\textbf{Step 2}} Select the maximum among these partial derivatives. Without loss of the generality, we suppose $|\frac{\partial \eta}{\partial \bm{P^{*}_{G}}}|$ to be the maximum. Reduce the corresponding deviation by half $\frac{1}{2}\Delta \bm{P^{*}_{G}}$.

\textit{\textbf{Step 3}} Solve the optimization model again with the updated deviation. Verify the satisfaction of the original stability CC. If not, return to \textit{\textbf{Step 1}}. The supplementary measure stops when all CCs are satisfied.

The intrinsic principle of the supplementary countermeasure is guaranteeing original probabilistic CCs by a more conservative strategy with sacrificing the economy to some extent. But, the measure is the most straightforward way to deal with exceptions caused by the linearization, which is convenient to implement and efficient. The effectiveness of the proposed supplementary countermeasure is demonstrated in detail in case study.
\vspace{-10pt}
\section{Case study}

\subsection{Simulation Settings}
\begin{figure}[htbp]
\centerline{\includegraphics[width=3in]{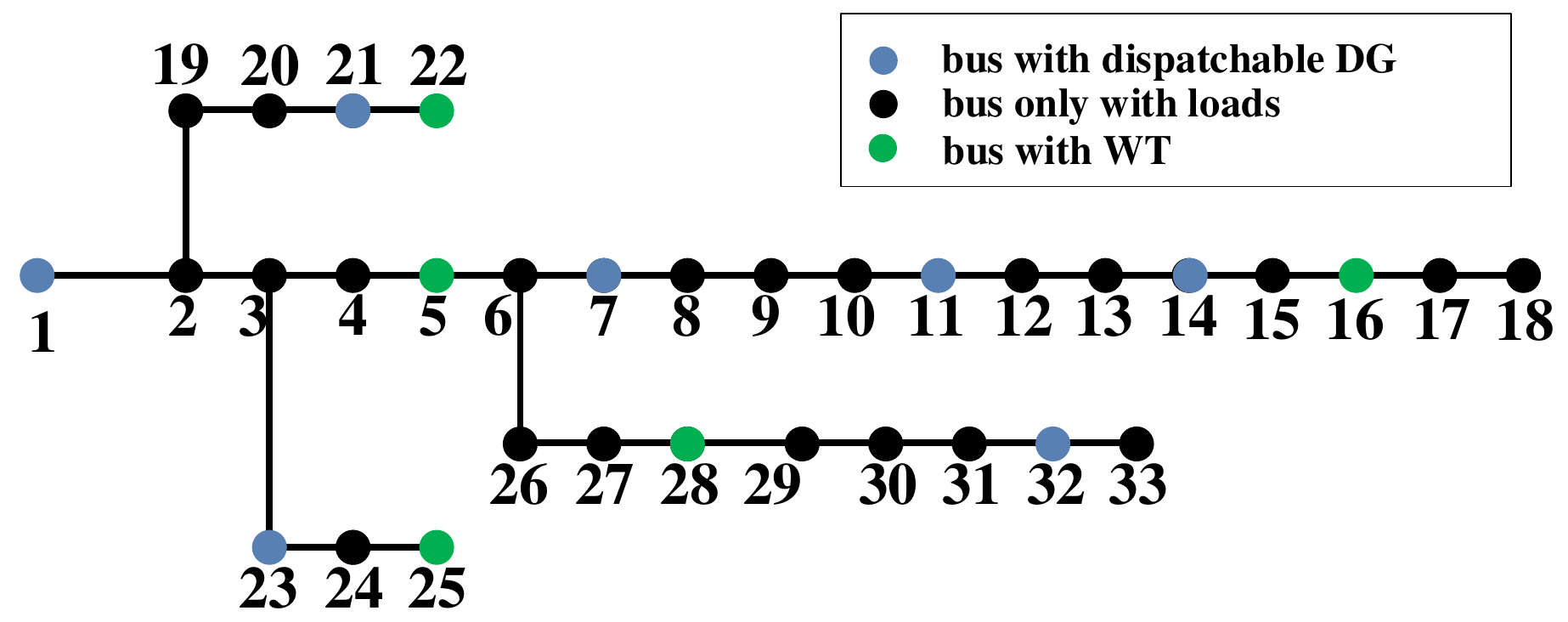}}
\caption{Diagram of modified IEEE 33 bus system}
\label{ieee33}
\end{figure}

We create a microgrid test model by inheriting the network parameters of IEEE 33-bus system, integrating droop-controlled DGs and disconnecting the slack bus (bus 1) with the subsection. We adopt this test model to
elaborate the proposed CC-SC-OPF model, as shown in Fig.~\ref{ieee33}. The original network parameters are same as MATPOWER package \cite{b28}. The following numerical tests are worked out on a computation platform with an Intel i7-10510U CPU processor and 16GB of memory. Different optimization models are all solved by MATLAB R2020a with the default setting.

\vspace{-10pt}
With considering the difference between microgrids and distribution networks, we reduce the power injection at each bus to 20\% of its original value. There are 7 dispatchable DGs at bus 1, 7, 11, 14, 21, 23, 32 with droop gains $K_{gpi}=1.3$, $K_{gqi}=7.8$. We set the measured frequency $F_{pi}=F_{qi}=20rad/s$ for each DG bus. The renewable DGs are WTs, which are located at bus 5, 16, 22, 25, 28. Their forecast errors are assumed to be less than 5\% by applying advanced forecast approaches. In addition, we set the upper bound of the stability index $\bar{\eta}=-0.15$, the confidence coefficients $\beta_{\eta}=0.05$, $\beta_G=\beta_V=0.01$. This paper considers the system operating process in one single timeslot, whose duration is 15 minutes.
\vspace{-10pt}
\subsection{Merits of the Proposed CC-SC-OPF Model}
To elaborate the performance of the proposed CC-SC-OPF model, we select the following four types of CC-OPF to compare their performances on stability and computational efficiency.

$\bullet \quad \textup{Case} \ 0$ : The original CC-OPF without considering the stability constraint.

$\bullet \quad \textup{Case} \ 1$ : The CC-SC-OPF which reformulates CCs into linear constraints with probability sensitivities given by Monte-Carlo simulation, as mentioned in \eqref{nup}.

$\bullet \quad \textup{Case} \ 2$ : The CC-SC-OPF which reformulates CCs into linear approximation forms between $(\eta, \bm{V}, \bm{P_G}, \bm{Q_G})$ and $\bm{\xi}$ where the stability sensitivity is obtained based on the numerical perturbation method.

$\bullet \quad \textup{Case} \ 3$ : The proposed CC-SC-OPF which reformulates CCs into linear approximation forms with analytical sensitivities. Compared to Case 2, this case analytically calculates the stability sensitivity which is introduced in \cite{b25}.

\subsubsection{Stability improvement}
\begin{figure}[htbp]
\centerline{\includegraphics[width=0.5\textwidth]{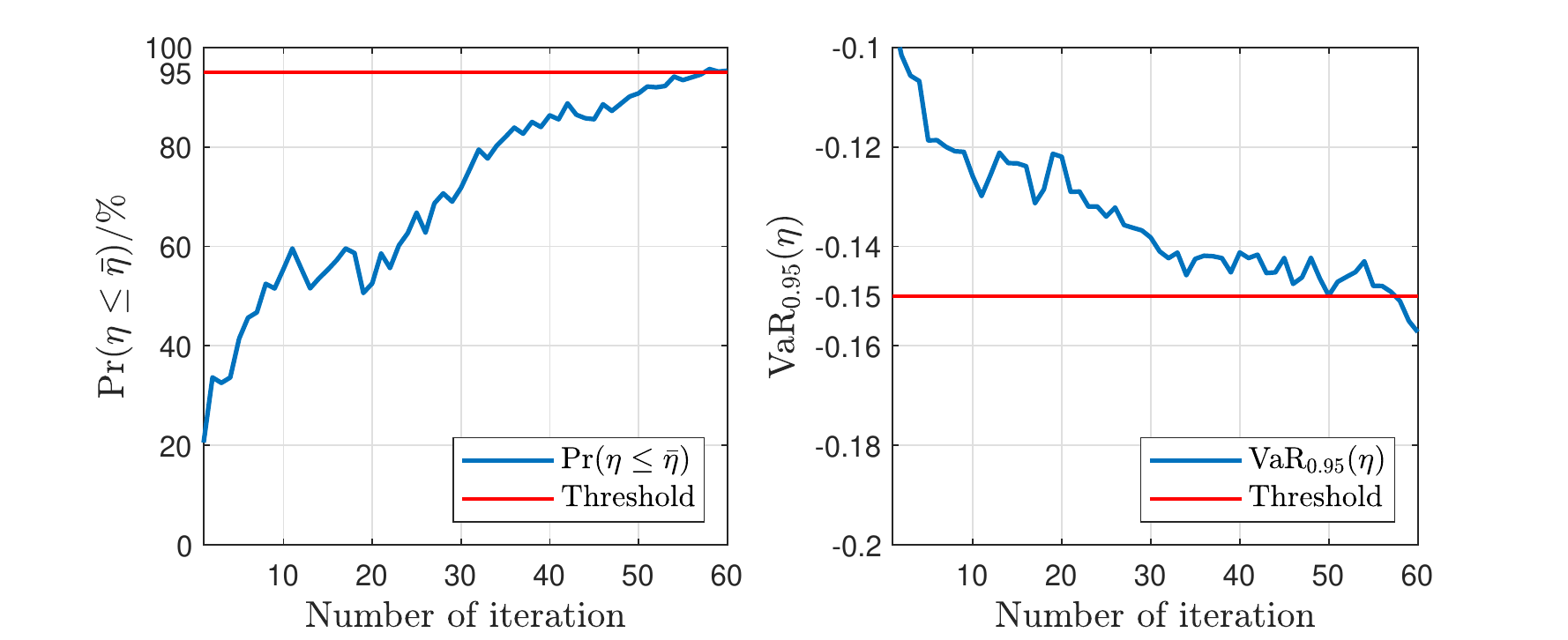}}
\caption{Stability improvement w.r.t the number of iterations}
\label{S3_iteration}
\end{figure}

This subsection mainly shows the stability improvement of the proposed model (Case 3) compared to the original case (Case 0). By adopting the proposed methodology, the optimization model can convergence after 58 iterations. The improvement of the system stability w.r.t. the number of iterations is elaborated in Fig.~\ref{S3_iteration}. It shows that the stability CC can be satisfied i.e., $\textup{Pr}(\eta \leq \bar{\eta}) \geq 95\%$ after 58 iterations. This CC is also equivalent to the VaR of the stability index being less than $\bar{\eta}=0.15$ when the confidence coefficient is 0.05.
\vspace{-10pt}

\vspace{-20pt}
\begin{figure}[htbp]
    \centering
    \subfigure[PDF of stability index before optimization]{
    \label{Eigenvalueprofile}
    \includegraphics[width=0.39\textwidth]{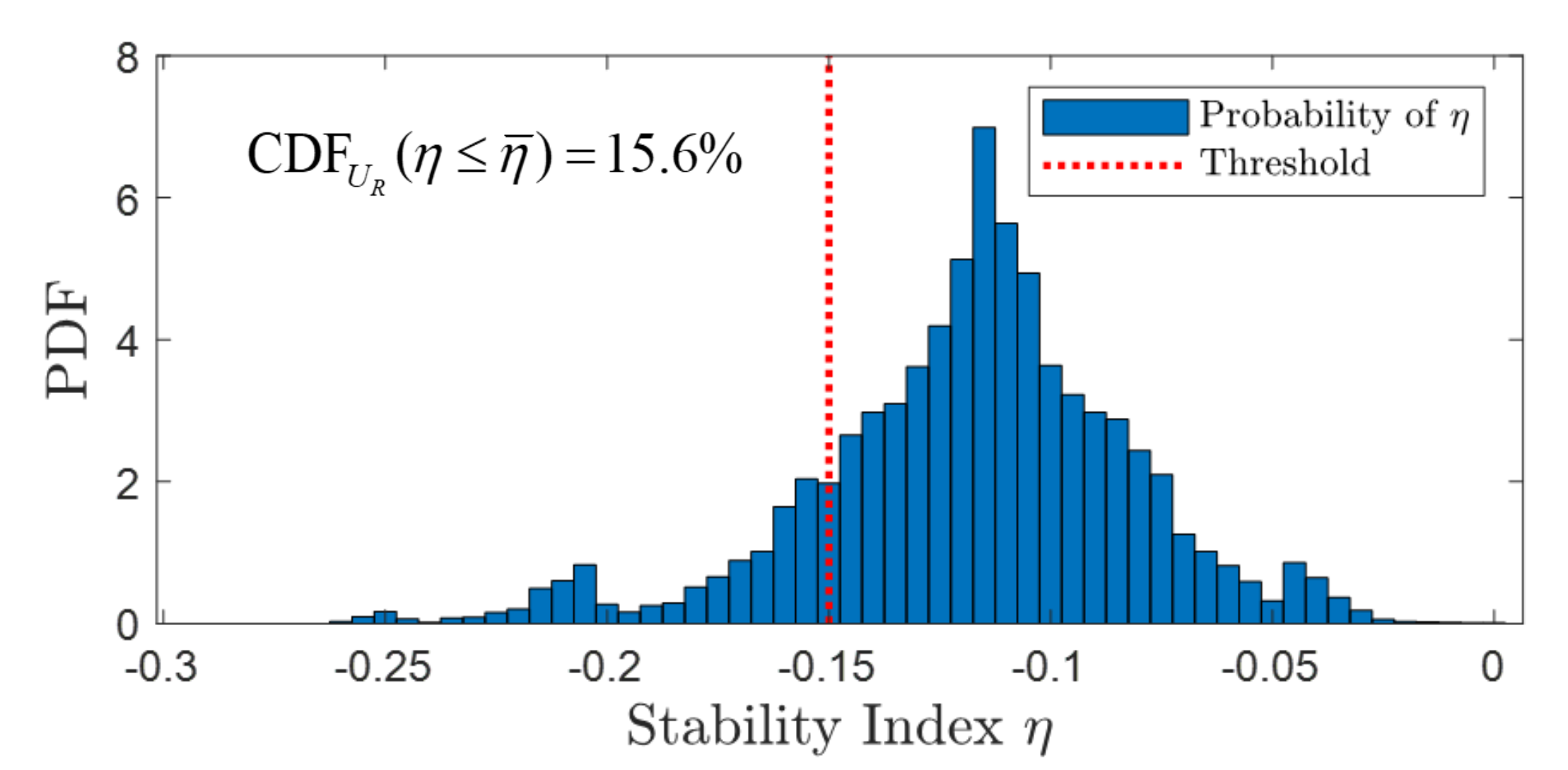}}
    \subfigure[PDF of stability index after optimization]{
    \label{response}
    \includegraphics[width=0.39\textwidth]{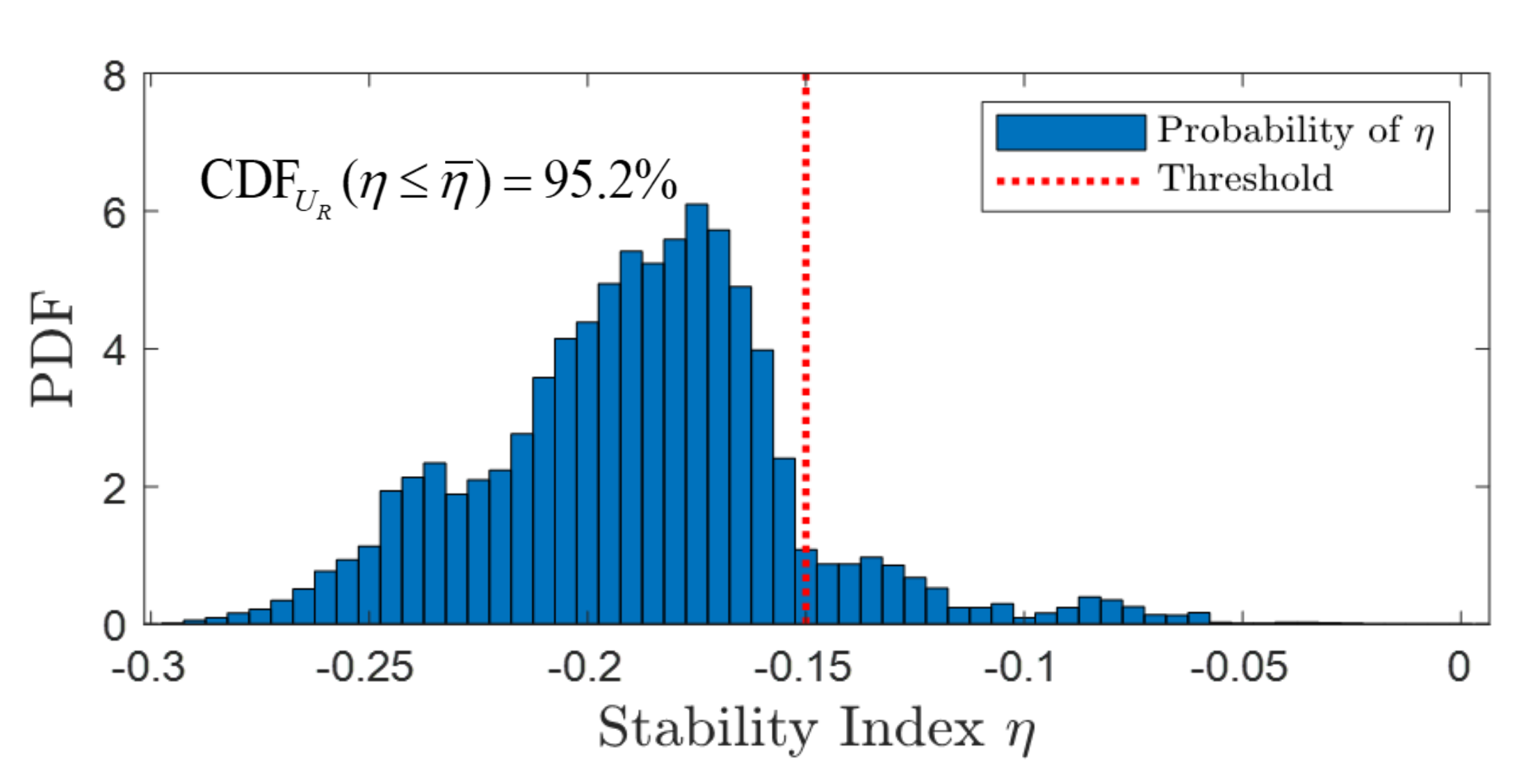}}
    \vspace{-5pt}
    \caption{PDF of stability index before and after optimization}
    \label{S1_stabilityindex_afterOPF}
\end{figure}
\vspace{-10pt}

With incorporating the stability CC into the CC-OPF model, the system stability can be significantly enhanced. Fig.~\ref{S1_stabilityindex_afterOPF} explicitly presents the original PDF of the stability index which is corresponding to the Case 0 and its optimized PDF corresponding to the Case 3. With the stability constraint, $\textup{CDF}_{\mU_R}(\eta \leq \bar{\eta})$ is managed to significantly increase and exceed $1-\beta_{\eta}$.


\subsubsection{Computational efficiency improvement}
Apart from the stability enhancement provided by the proposed CC-SC-OPF model, the computational efficiency is also dramatically improved by adopting the proper linearization as well as the analytical sensitivity. This subsection mainly shows the computational efficiency improvement through comparing the CPU time of different cases (Case 1, Case 2 and Case 3).

Besides, to investigate how the uncertainty affects the efficiency performance, we define the uncertainty degree of WT output noted by $\mu_W$,
\begin{equation}
\mu_W = \frac{1}{n_w}\sum_{i=1}^{n_w}\frac{\epsilon_{w,i}}{P^{pre}_{w,i}}
\end{equation}
where $n_w$ denotes the number of WTs integrated in the microgrid, $\epsilon_{w,i}$ denotes the forecast error and $P^{pre}_{w,i}$ is the predicted output of the WT.

\begin{figure}[htbp]
\centerline{\includegraphics[width=0.5\textwidth]{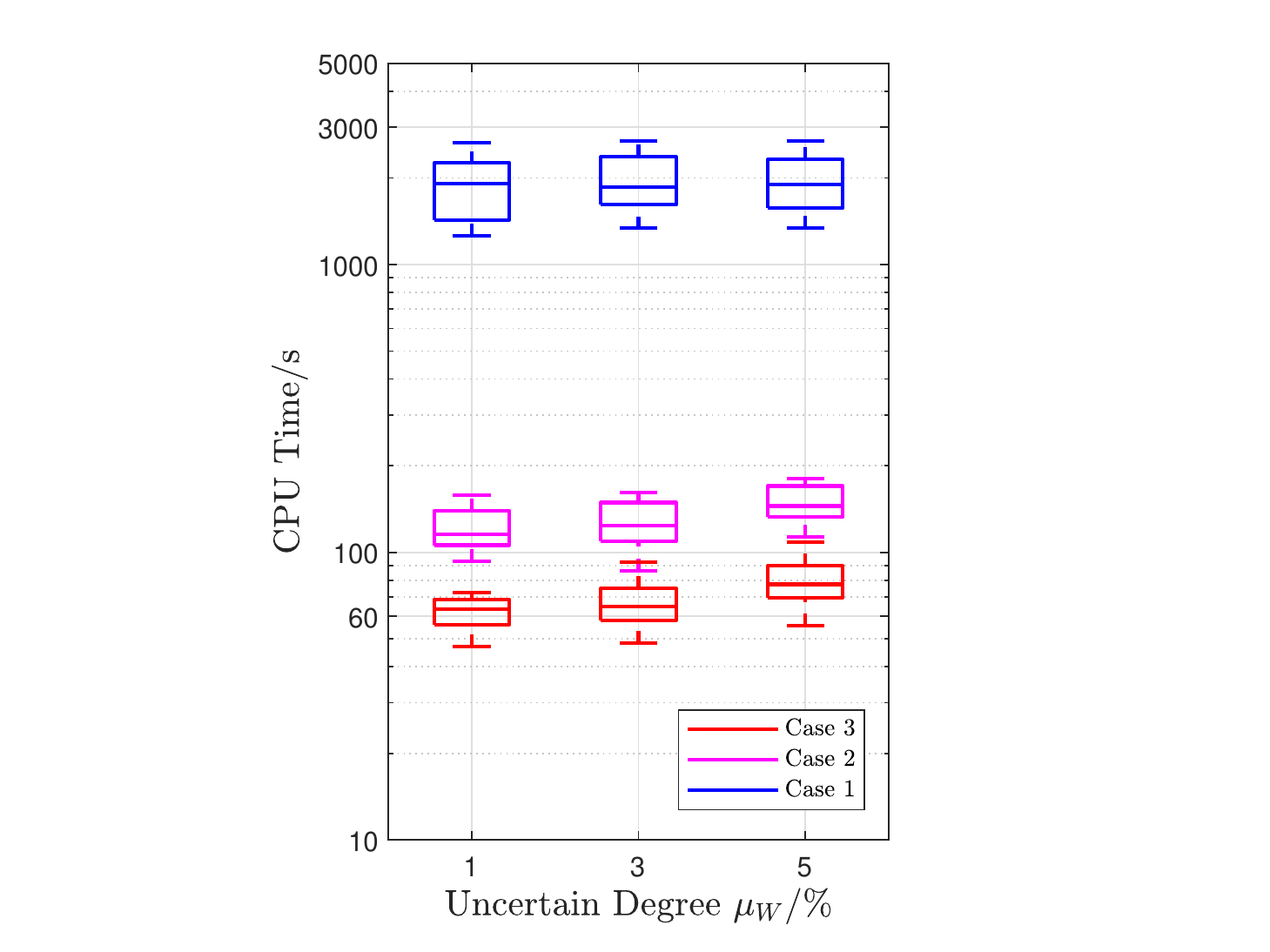}}
\caption{CPU time under different cases}
\label{CPUtimeunderthreecases}
\end{figure}

To obtain more numerical tests for verifying the computational efficiency, we choose different values of multiplier $r_g$ from 50\% to 150\%, and replace original DG droop gains $K_{gpi}$ and $K_{gqi}$ by $r_gK_{gpi}$ and $r_gK_{gqi}$. The average and maximal CPU time for solving Case 1 to Case 3 with different uncertainty degrees are provided in Table.~\ref{differentOPF}. Moreover, Fig.~\ref{CPUtimeunderthreecases} gives the graphical illustration of detailed computational improvements.
\begin{table}[htbp]
\vspace{-5pt}
\caption{Average and maximal CPU times under different cases}
\vspace{-10pt}
\begin{center}
    \label{differentOPF}
    \begin{threeparttable} 
        \begin{tabular}{cccccccccc} \toprule
            \multirow{2}{*}{} & \multicolumn{2}{c}{Case 1} & \multicolumn{2}{c}{Case 2} & \multicolumn{2}{c}{Case 3}\\ \cmidrule(r){2-3} \cmidrule(r){4-5} \cmidrule(r){6-7}
            & $T_\textup{avg}$\tnote{1} & $T_{\max}\tnote{2}$ & $T_\textup{avg}$ & $T_{\max}$ & $T_\textup{avg}$ & $T_{\max}$ \\ \hline

            $\mu_W \approx 1\%$ & 1933 & 2616 & 116.3  & 149.5  & 61.01  & 77.64 \\
            $\mu_W \approx 3 \%$ & 2068 & 2738 & 127.6  & 159.9  & 60.75  & 85.83 \\
            $\mu_W \approx 5 \%$ & 1966 & 2715 & 149.5  & 171.1  &83.21   & 101.3 \\
            \bottomrule
        \end{tabular}
        \begin{tablenotes}
            \item [1] $T_\textup{avg}$ denotes the average CPU time in seconds.
            \item [2] $T_{\max}$ denotes the maximal CPU time in seconds.
        \end{tablenotes}
    \end{threeparttable}
\end{center}
\vspace{-15pt}
\end{table}

Since Case 1 directly adopts the numerical sensitivity of probabilities, in each iteration it needs two rounds of Monte-Carlo Simulations, which are extremely time consuming. From Table.~\ref{differentOPF}, we observe that the CPU time of Case 1 is around 30 times longer than the CPU time of our proposed model (Case 3). Consequently, the applied linearization approach can significantly improve the computational efficiency.
\vspace{-10pt}
\vspace{-10pt}

Although both Case 2 and Case 3 adopt the linearization to help reformulate original CCs, the computational efficiency of two cases are entirely different. The CPU time of Case 2 is twice as longer as Case 3. This improvement is brought by the analytical sensitivity, which is by-product in the solution process and can be obtained without additional computations \cite{b25}. While the numerical perturbation based sensitivity needs twice eigen-analysis. Although the numerical perturbation-based approach is easy to implement, it is inaccurate because the result strongly depends on the perturbation value. Instead of adopting the numerical approach, this paper uses an analytical sensitivity analysis to accurately calculate the partial derivative of the stability index to its decision variables. Accordingly, apart from the improvement on the computational efficiency, the accuracy will be significantly improved.
\vspace{-10pt}
\begin{figure}[htbp]
\centerline{\includegraphics[width=0.39\textwidth]{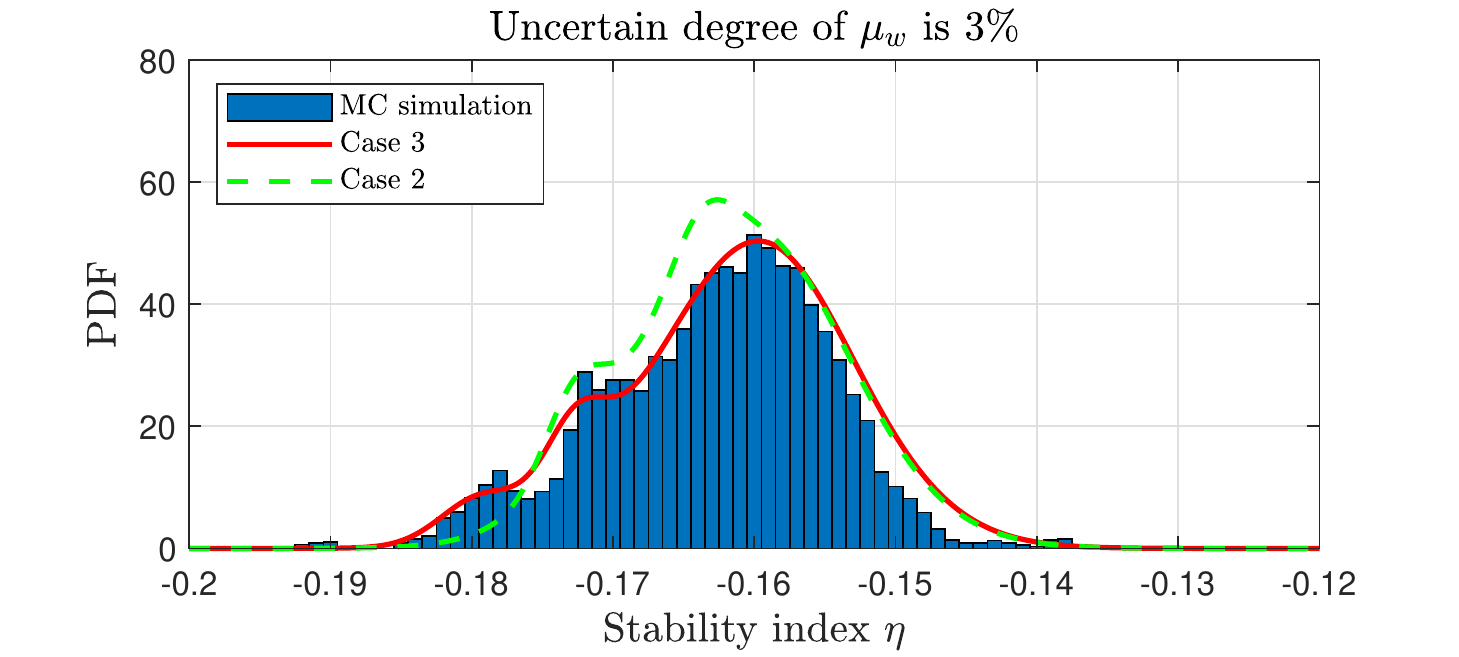}}
\caption{Graphically illustration of accuracy improvement}
\label{sensitivityaccuracy}
\end{figure}
\vspace{-10pt}

Fig.~\ref{sensitivityaccuracy} presents that Case 3 (red curve) estimates the PDF better when $\mu_w$ is set to be $3\%$. The error index called ``Average Root Mean Square (ARMS)'' is adopted to quantify the accuracy of different cases. The smaller ARMS means the higher correctness. Table.~\ref{ARMSsensitivity} quantitatively reveals the accuracy improvement by employing the analytical stability sensitivity. Compared to Case 2, the analytical sensitivity brings more than 10\% ARMS reduction in CDF and PDF estimations.
\begin{table}[htbp]
\vspace{-5pt}
\caption{Accuracy improvement through analytical stability sensitivity}
\vspace{-10pt}
\begin{center}
    \label{ARMSsensitivity}
    \renewcommand\tabcolsep{3.8pt} 
    \begin{threeparttable} 
        \begin{tabular}{cccc} \toprule
            & $\mu_W \approx 1\%$ & $\mu_W \approx 3\%$ & $\mu_W \approx 5\%$\\
            \hline
            $\textup{ARMS}_\textup{red}$(CDF;PDF) & 13.3\%; 10.5\% & 16.6\%; 10.6\%  & 18.7\%; 10.3\% \\
            \bottomrule
        \end{tabular}
    \end{threeparttable}
\end{center}
\vspace{-15pt}
\end{table}

\vspace{-10pt}





\subsection{The accuracy of the linear approximation of stability CC}
This subsection presents the accuracy of the linear approximation for the stability CC and GMM estimation for non-Gaussian uncertainty. As mentioned in previous section, the mappings from the DG active/reactive outputs, the bus voltage to are approximately linear. While, the mapping from the stability index to $\{P^{*}_{G,i}, Q^{*}_{G,i}, V^*_{G,i}, \epsilon_{R}\}$ is highly nonlinear. Compared to the stability index, the proposed model can more accurately implement the probabilistic DG active/reactive outputs and bus voltage. Accordingly, this subsection mainly focuses on the accuracy to describe the probabilistic stability index.

The proposed model adopts the GMM to estimate the RES forecast error $\epsilon_{R}$ which follows a non-Gaussian distribution. This part presents results that demonstrate the GMM can efficiently fit the PDF of stability index. The proposed model is compared with following cases, and we choose the uncertain degree to be $1\%$, $3\%$ and $5\%$.\\
$\bullet \quad \textup{Case} \ 4$ : The CC-SC-OPF applying the method based on Gaussian assumption for RES forecast error.\\
$\bullet \quad \textup{Case} \ 5$ : The CC-SC-OPF applying Gram-Charlier (GC) expansion method \cite{b60} to estimate the PDF.\\
$\bullet \quad \textup{Case} \ 6$ : The CC-SC-OPF applying Cornish-Fisher (CF) expansion method \cite{b61} to estimate the PDF.

\begin{table}[htbp]
\vspace{-5pt}
\caption{Accuracy and efficiency performance of different probability computing methods}
\vspace{-10pt}
\begin{center}
    \label{ARMS}
    \renewcommand\tabcolsep{3.5pt} 
    \begin{threeparttable} 
        \begin{tabular}{cccccccccc} \toprule
            \multirow{2}{*}{} & \multicolumn{2}{c}{Case 4} & \multicolumn{2}{c}{Case 5} & \multicolumn{2}{c}{Case 6}\\ \cmidrule(r){2-3} \cmidrule(r){4-5} \cmidrule(r){6-7}
            & $\textup{ARMS}_\textup{red}$\tnote{1} & $T\tnote{2}$ & $\textup{ARMS}_\textup{red}$ & $T$ & $\textup{ARMS}_\textup{red}$ & $T$ \\
            & (CDF; PDF) &  & (PDF) &  & (CDF) &  \\
            \hline

            $\mu_W \approx 1\%$ & 76.8\%; 50.1\% & 11.6 & 37.6\%  & 265.2  & 45.7\%  & 277.3 \\
            $\mu_W \approx 3\%$ & 80.3\%; 42.5\% & 12.3 & 41.5\%  & 259.3  & 51.6\%  & 285.8 \\
            $\mu_W \approx 5\%$ & 81.1\%; 41.9\% & 10.9 & 39.1\%  & 272.9  & 56.2\%  & 301.3 \\
            \bottomrule
        \end{tabular}
        \begin{tablenotes}
            \item [1] $\textup{ARMS}_\textup{red}$ denotes the ARMS reduction w.r.t. Case 4, Case 5 and Case 6, respectively.
            \item [2] $T$ denotes the CPU time in seconds.
        \end{tablenotes}
    \end{threeparttable}
\end{center}
\vspace{-15pt}
\end{table}

\begin{figure}[htbp]
\centerline{\includegraphics[width=0.39\textwidth]{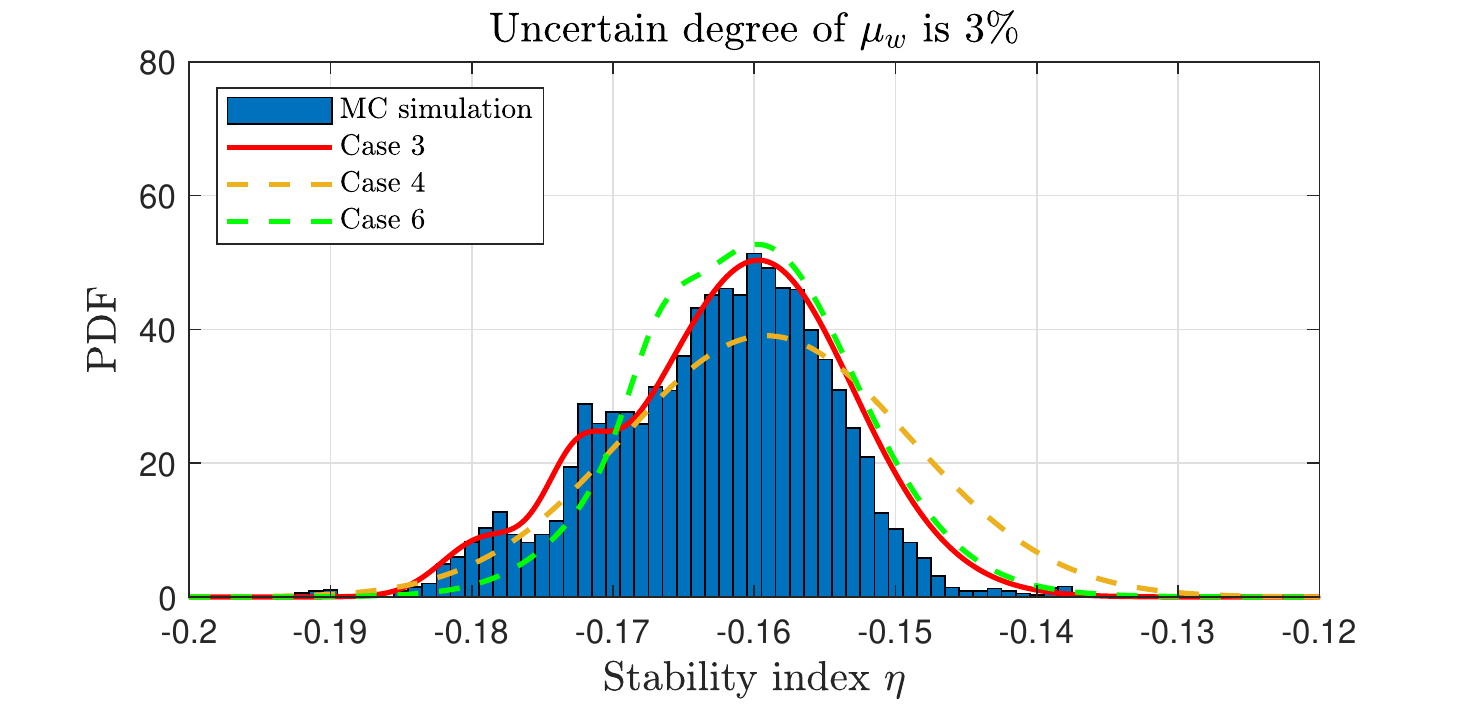}}
\caption{Different methods to estimate PDF of stability index}
\label{GMMaccuracy}
\end{figure}

Table.~\ref{ARMS} reveals that GMM can significantly reduce the ARMS w.r.t. Gaussian assumption, GC and CF expansion methods. Take the PDF of the stability index as an example to give the validation graphically. The uncertain degree is set to be $3\%$. Fig.~\ref{GMMaccuracy} illustrates that the method based on Gaussian assumption (yellow dash curve) is not accurate. It underestimates the probability in the head/middle region and overestimates the probability in the tail. Compared to Gaussian assumption, the GC expansion method (green dash curve) improves the accuracy, but it still underestimates the PDF in the head region. Moreover, the GC and CF methods are also time consuming. This subsection clearly validates the accuracy and effectiveness of GMM in dealing with non-Gaussian uncertainty.
\vspace{-20pt}
\vspace{-25pt}
\subsection{Effectiveness of Supplementary Corrective Countermeasure}
Since the proposed CC-SC-OPF model applies the linearization approach to reconstruct highly nonlinear probabilistic CCs, a few exceptions in solutions inevitably appear. To deal with this problem, this paper uses the supplementary countermeasure to correct. Take the PDF of stability index as an instance to present the effectiveness of this countermeasure.
\vspace{-10pt}
\begin{figure}[htbp]
    \centering
    \subfigure[PDF of stability index before optimization]{
    \label{Eigenvalueprofile}
    \includegraphics[width=0.39\textwidth]{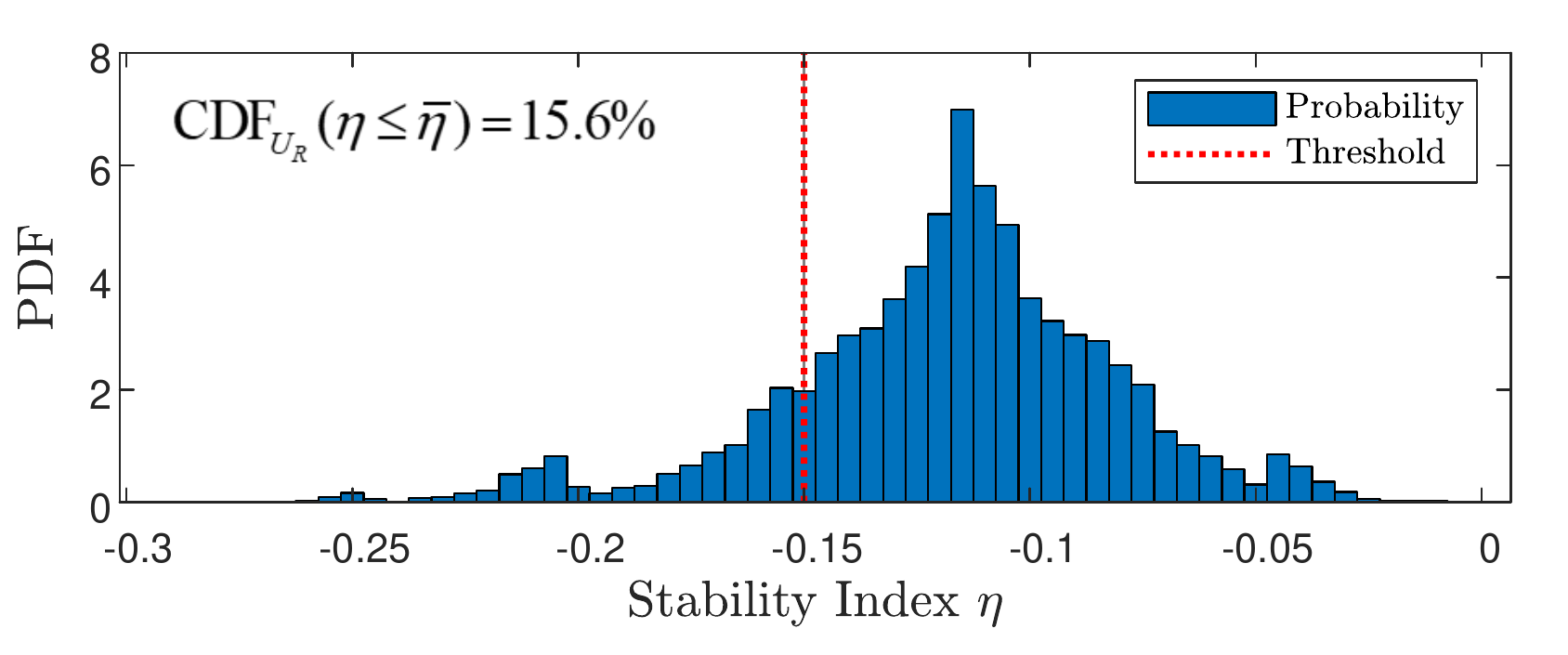}}
    \subfigure[Improper PDF of stability index after optimization]{
    \label{response}
    \includegraphics[width=0.39\textwidth]{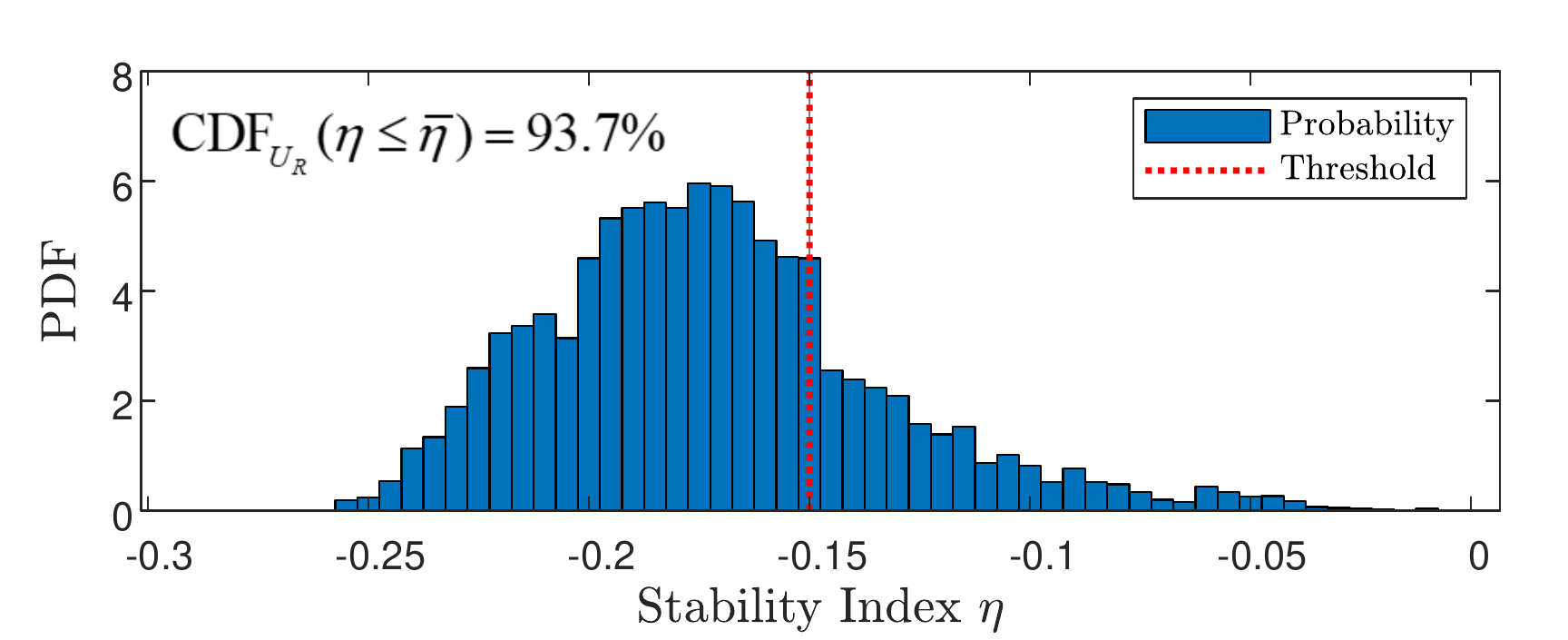}}
    \subfigure[Proper PDF of stability index by using the countermeasure]{
    \label{response}
    \includegraphics[width=0.39\textwidth]{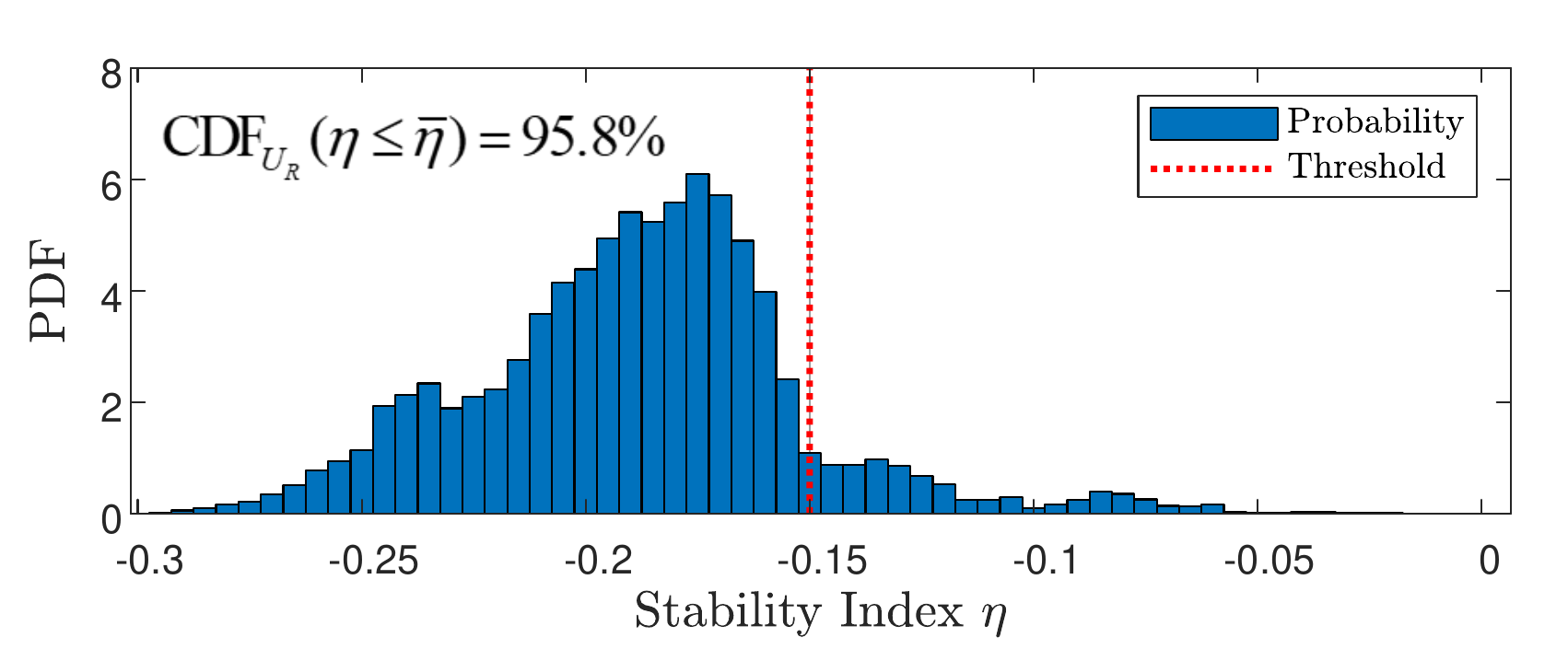}}
   \vspace{-5pt}
    \caption{The optimization procedure of stability index PDF}
    \label{stabilityindexpdfwithOPF}
\end{figure}
\vspace{-10pt}

Fig.~\ref{stabilityindexpdfwithOPF} shows how the supplementary countermeasure works in the optimization procedure when an exception appears. Owing to the error caused by the linearization, the inproper PDF is obtained after solving the optimization model (the middle figure). Nevertheless, this solution cannot satisfy the original stability CC, i.e. $\textup{CDF}_{\mU_R}(\eta \leq \bar{\eta})$ is smaller than 95\%. Under this circumstance, the supplementary countermeasure is activated. After reducing by half the deviation corresponding to the maximal partial derivative in the last iteration, the corrective PDF is recalculated. The proper PDF is presented in the bottom figure, which well satisfies the CC with $\textup{CDF}_{\mU_R}(\eta \leq \bar{\eta}) = 95.8\%$. Consequently, simulation results clearly verify that this easy to implement measure can efficiently correct these exceptions without additional computations in the optimization procedure.



\section{Conclusion}
This paper has proposed a CC-SC-OPF model with incorporating GMM-based estimations to cope with small-signal stability issues brought by RES non-Gaussian uncertainties in isolated microgrids. To guarantee the probabilistic small-signal stability without increasing the total cost, a bi-level CCO framework is established. The upper level focuses on minimizing the expected cost with considering the stability CC. The lower SDP problem aims to calculate the stability index by applying the Lyapunov equation. By applying GMM to fit RES non-Gaussian forecast errors, analytical sensitivity cuts are designed to linearly approximate the stability index and other operational variables and reconstruct highly nonlinear CCs. The bi-level OPF model with reformulated CCs is solved by Benders decomposition based-methodology with the supplementary corrective measure. Simulation results on the 33-bus microgrid reveal that compared to other traditional approaches, such as, Monte Carlo simulation, the proposed model has the capability of converging in much shorter CPU time with adopting the sensitivity cuts. This benefit enables the significant stability improvement in every single timeslot (15 minutes).




\ifCLASSOPTIONcaptionsoff
  \newpage
\fi

{\footnotesize
\bibliographystyle{IEEEtran}

}

\end{document}